\documentclass[a4j]{article}
\usepackage[psamsfonts]{amssymb}
\usepackage{amsmath}
\usepackage{amsfonts}
\usepackage{fancybox}
\usepackage{fancyhdr}
\usepackage{bm}
\usepackage[dvipdfm]{graphicx}
\makeatletter
\long\def\@makefntext#1{\parindent 1em\noindent
  \hbox to 2em{\hss$^{\@thefnmark}$~}%
  \@tempdima\columnwidth\advance\@tempdima-2em
  \parbox[t]{\@tempdima}{#1}}
\makeatother
\begin{document}
\title{Learning curve for collective behavior of zero-intelligence agents in successive job-hunting processes 
with a diversity of Jaynes-Shannon's MaxEnt principle}

\author{
He Chen\footnote{Email: chen@complex.ist.hokudai.ac.jp} \, and \,Jun-ichi Inoue\footnote{Email: jinoue@cb4.so-net.ne.jp}
\mbox{}\\
Graduate School of Information Science and Technology \\
Hokkaido University, N14-W9, Kita-ku, Sapporo 060-0814, Japan}
\date{\today \footnote{
This manuscript was accepted by  {\it Evolutionary and Institutional Economics Review}.}}
\maketitle
\begin{abstract}
Inspired by the unsupervised  learning or self-organization in the machine learning context, 
here we attempt to draw `learning curve' for the collective 
behavior of job-seeking `zero-intelligence' labors in 
successive job-hunting processes. 
Our labor market is supposed to be opened especially for university graduates in Japan, 
where the students have several successive chances $n=0,1,2,\cdots$ to obtain their positions  
within an academic (business) year. 
In this sense,  
the `cumulative unemployment rate' in our model system is 
regarded as an error-measurement in the collective intelligence of students,  
and the job-hunting stage $n$-dependence of the error constructs a learning curve. 
In our simple toy-model of 
probabilistic labor market, the diversity of students' behavior 
is built-in by means of the Jaynes-Shannon's MaxEnt (Maximum Entropy) principle. 
Then, we discuss the speed of convergence for the error-measurement, where 
we consider a scenario in which 
the students do not use any information about the result of job-hunting processes in the previous stage. 
Our approach enables us to examine the existence of the condition on which macroscopic quantity, say, 
`stage-wise unemployment rate' becomes  `scale-invariant' in the sense that 
it does not depend on the job-hunting stage $n$. 
From the macroscopic view point, 
the problem could be regarded as a human resource allocation. 
\end{abstract}
\mbox{}\\
{\bf Keywords:} Labor market, Microscopic description of macroeconomics, Learning curve, 
Statistical mechanics, Jaynes-Shannon's MaxEnt principle, Agent-based modeling, Econophysics
\mbox{}\\
{\bf JEL Codes:} J42, E24, C51
\section{Introduction}
\label{sec:Intro}
Deterioration of the employment rate is now one of the most worrying problems in Japan 
\cite{keizaisanngyou,kouseiroudou,works}. Actually, various attempts to overcome these difficulties 
have been done by governments. 
Apparently, 
labor (work) is important not only for each of us 
to earn our daily bread, 
but also for our state to 
keep the revenues 
by collecting the taxes from labors. 
Especially, 
in recent Japan, 
the employment rate is getting worse and 
the government 
has over-issued quite a lot of government bonds  
to compensate a lack of the tax revenues 
and the national debt is now 
becoming a serious risk to cause a national-wide bankruptcy.

To consider the effective policy and to carry out it 
for sweeping away the unemployment uncertainty caused by 
the so-called `mismatch', 
it seems that we should investigate the labor markets 
scientifically and if possible, one should simulate 
artificial labor markets in computer to reveal 
the essential features of the problem. 
In fact, in macroeconomics (labor science), there exist a lot of effective attempts to discuss 
the macroscopic properties \cite{Aoki,Boeri,Roberto,Fagiolo,Neugart}. 
However, apparently, the macroscopic approaches 
lack of their microscopic view points, namely, in their arguments, 
the behavior of microscopic heterogeneous agents 
such as labors or companies are neglected. 

In our preliminary studies \cite{Hino}, we attempted to construct such microscopic models  
of probabilistic labor markets in which the labors (job-seekers) and the companies behave 
probabilistically. Then, we evaluated 
the unemployment and the inflation rates 
as relevant macroscopic quantities and 
draw the so-called Philips curves \cite{Philips}, which have a universal fitting curve: 
$\mbox{(inflation rate)} = \mbox{(unemployment rate)}^{c}+b,\,b,c<0$. 
However, in the model, we introduced a lot of parameters 
to describe the labor market and 
it seems not to be far-sighted to reveal the 
essential properties of the market and 
we need much more simpler or `minimal' 
models to discuss the macroscopic behaviour from the 
microscopic descriptions. 

Taking this fact in mind, in the references \cite{Chen,Chenb} 
we proposed a simple probabilistic model 
based on the concept of statistical mechanics  
for probabilistic labor markets, 
in particular, Japanese labor markets for 
university (college) graduates. 
In this agent-based modeling, 
we introduced the energy function for each company. 
The energy function 
describes the power of each company to 
gather the entry sheets (application letters) of the students. 
The energy consists of 
two distinct factors, 
namely, ranking preference and 
market history effects. 
The later acts as `negative feedback' on the 
ranking preference.  
Then, we evaluated the relevant macroscopic 
quantity, namely, 
unemployment rate by means of the microscopic variables. 
We also drew the so-called 
Beveridge curve (UV curve: `Unemployment -Vacancy') which 
shows the relationship 
between the unemployment rate and 
the number of vacancies (job-offer ratio). 
We found that 
the system undergoes a 
phase transition with spontaneous symmetry breaking 
from the good employment rate phase to 
the poor employment rate phase 
when one increases the degree of 
ranking preference. 
In the references \cite{Chen,Chenb}, 
we also utilized  
the macroscopic Neugart model \cite{Neugart} 
which describes the Philips curve as 
an attractor of chaotic maps. 
We used the update equation for 
the inflation rate accompanying with 
our probabilistic model to derive the 
Philips curve for our model system. 
We found that the Philips curve is actually obtained 
by the `mixture model'.

In Japanese labor markets for university graduates, 
the students have several successive chances to obtain their positions within an academic (business) year.
After each opportunity (stage) is over, the numbers of both students and positions 
(namely, the so-called `effective system size')  
are reduced remarkably because several excellent students get their job immediately after they apply for. 
Accordingly, the macroscopic quantities such as 
unemployment rate $U$, labor shortage ratio $\Omega$ or 
job offer ratio $\alpha$ 
changes as $U^{(0)} \to \cdots U^{(n)}$, 
$\Omega^{(0)} \to \cdots \Omega^{(n)}$ or 
$\alpha^{(0)} \to \cdots \alpha^{(n)}$. 
However, it is highly non-trivial whether some of them 
could be scale invariant or not, 
that is to say, 
it is not easy for us to conclude that there exists the condition on which a macroscopic quantity, say, 
$U^{(n)}$ satisfies $U^{(n)}=\mbox{\it constant}$ for any job-hunting stage $n$. 
Here we shall focus on this issue and clarify the $n$-dependence (or independence) of Japanese 
labor markets for university graduates by making use of both extensive 
computer simulations and empirical data analysis. 
In order to 
investigate these aspects of labor markets, 
here we attempt to draw `learning curve' for the collective 
behavior of job-seeking `zero-intelligence' labors in 
successive job-hunting process. 
Our labor market is supposed to be opened especially for university graduates in Japan, 
where the students have several successive chances $n=0,1,2,\cdots$ to obtain their positions  
within an academic (business) year. 
In this sense,  
the `cumulative unemployment rate' in our model system is 
regarded as an error-measurement in the collective intelligence of students 
and the job-hunting stage $n$-dependence of the error constructs a learning curve. 
In our toy-model of 
probabilistic labor market, the diversity of students' behavior 
is built-in by means of the Jaynes-Shannon's MaxEnt (Maximum Entropy) principle. 
We discuss the speed of convergence for the error measurement, where 
we consider a scenario in which 
the students do not use any information about the result of job-hunting processes in the previous stages. 

This paper is organized as follows. 
In the next section \ref{sec:evidence}, 
we introduce several important concepts used in this paper. 
In particular, we account for two distinct quantities, 
namely, `cumulative' and `stage-wise' quantities. 
The generic properties of them would be explained.  
In the same section, we also show several important empirical findings. 
In section \ref{sec:model}, 
we introduce our probabilistic labor market which is constructed 
by means of Jaynes-Shannon's MaxEnt principle. 
The numerical results are given in the next section \ref{sec:result}. 
The last section is devoted to summary. 
\section{Empirical evidence}
\label{sec:evidence}
\begin{figure}[!htb]
\resizebox{\textwidth}{!}{%
  \includegraphics{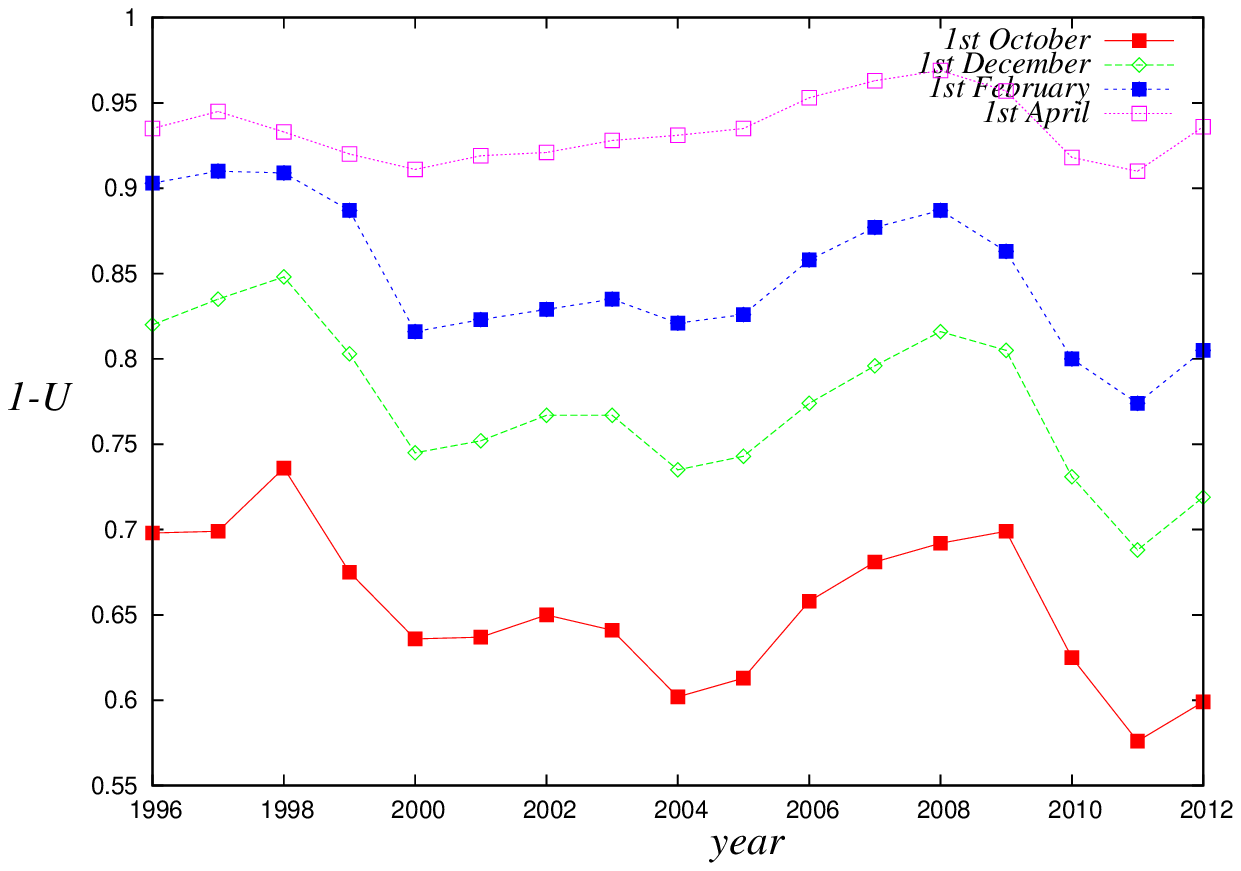}
 \includegraphics{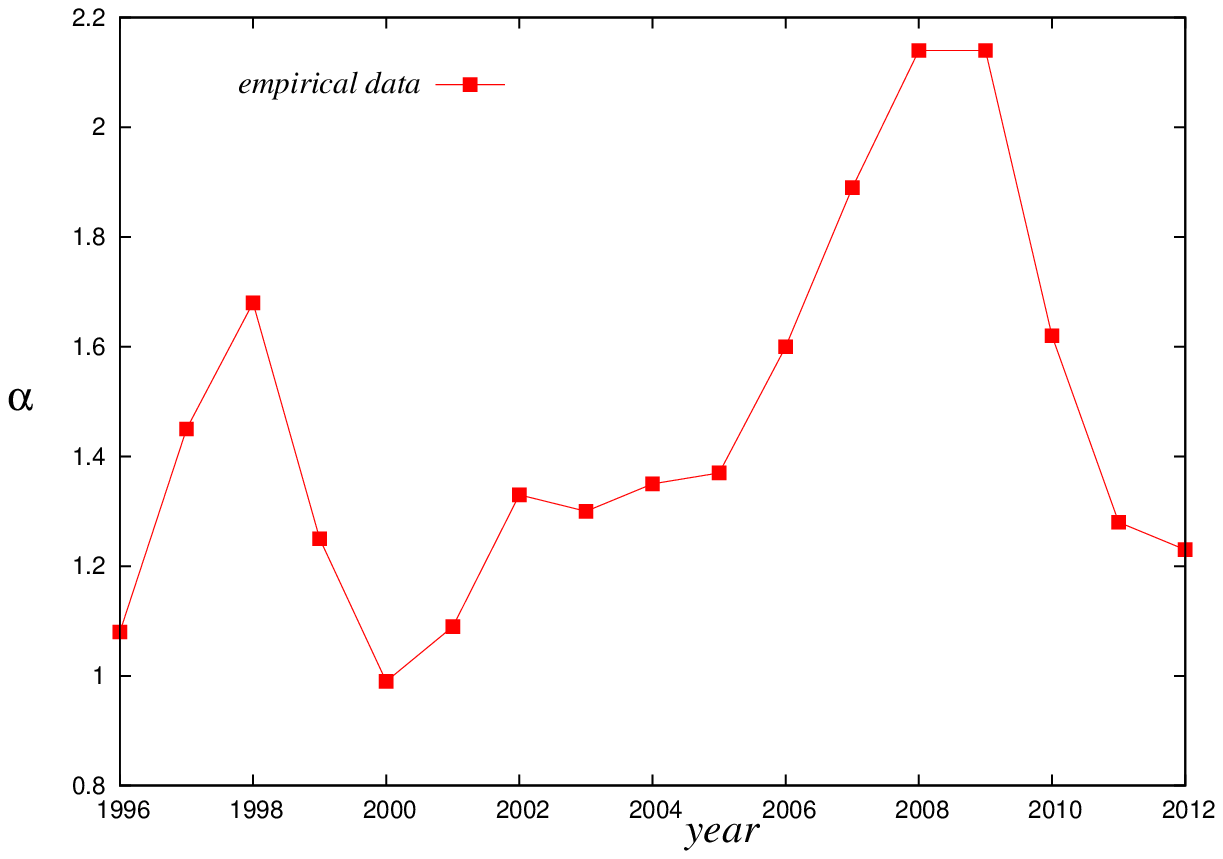}
}
\caption{\footnotesize 
The left panel shows the changing trends in cumulative employment rate for each employment opportunities, 
based on employment statistical data of university graduates in 1996 to 2012 by MEXT (Ministry of Education, 
Culture, Sports, Science and Technology) \& MHLW 
(Ministry of Health, Labor and Welfare) \cite{mext}. 
The right panel exhibits 
the annual job-offer ratio at 
October, that is,  $\alpha^{(0)} \equiv \alpha$. 
We find that 
the $\alpha$ is greater than unity except for in 2000. 
}
\label{fig:fg2}
\end{figure}

As we already mentioned in the previous section, 
employment uncertainty 
in young generation, especially in 
university graduates  
has been one of the serious social problems in Japan. 
Actually, 
the employment rate 
(the informal acceptance rate) reported in four times (October, December and February, April in the next year) within a business year is getting worse. 
In Figure \ref{fig:fg2} (left), 
we plot the empirical trends of employment rate $1-U$ changing in year 
for the four employment opportunities (let us specify the stage by 
$n$ as $n=0,1,2$ and $3$) in the last  seventeen years. 
From this panel, 
we clearly observe that 
in each year the employment rate increases 
monotonically as the stage $n$ goes on (see the right panel). 
This means that 
the students and companies make up these four labor markets having 
the different (reduced) system sizes. 
Namely, 
after the first $n=0$ stage, 
several excellent students 
go away from the market and 
several popular companies also close recruiting students 
because they might satisfy their quota immediately after they start to advertise for 
the candidates for the posts. Hence, the size of the market 
(the number of students and positions) 
is definitely getting small as the job-hunting stage goes on as $n=0,1,2,\cdots$. 

In the right panel of Figure \ref{fig:fg2}, 
we plot the annual job-offer ratio $\alpha$ 
which is usually released to the public in every October by 
{\it MEXT} (Ministry of Education, 
Culture, Sports, Science and Technology) \& {\it MHLW}  
(Ministry of Health, Labor and Welfare) \cite{mext}.  
From this panel, we observe that the $\alpha$ is greater than $1$ 
except for the statistics in 2000. 
In this sense, Japanese labor markets for university graduates 
are almost `seller's market' in which the number of 
opened positions is larger than that of job-seeking students. 
\subsection{`Stage-wise' unemployment rate and `cumulative' employment rate}
In the previous section, 
we found that the employment rate $1-U$ is getting better as the term goes on as 
October, December, February, and April. 
This fact is naturally accepted because 
the employment rate $1-U$ is `cumulative quantity'. 
In this sense, the $1-U$ is a monotonically increasing function of the term. 
However, when we assume that the market might shrink gradually 
as the term goes on, namely, 
the size of the market such as 
the number of students, positions might decrease , 
one could define the `size-reduced market' which is 
induced from the market at the previous stage (term) after. 
In this sense, there might exist successive 
opportunities for students who failed in the previous stages 
to apply to the positions in companies 
in the size-reduced market after each term is over.  

Then, an interesting query might arise in our mind. Namely, 
is it possible for us to conclude that the unemployment rate at the stage $n$ --- 
what we call `stage-wise' unemployment rate --- should 
be the size (scale)-independent as $U^{(n)}=U (=\mbox{\it constant})$?, where we defined 
the following 
`stage-wise' unemployment rate 
\begin{equation}
U^{(n)}=
\frac{\mbox{$\#$ of students who do not get any job even after the $n$-stage is over}}{
\mbox{$\#$ of students who are looking for the job  
after the $(n-1)$-stage is over}}. 
\end{equation}
To answer this highly relevant question, 
let us define 
the `cumulative' employment rate at each stage $n=1,2, \cdots$ by 
\begin{equation}
(1-U)_{n} = 
\frac{\mbox{total $\#$ of students who got the position until the stage $n$}}
{\mbox{total $\#$ of students}}.
\end{equation}
Obviously, 
unemployment rate 
shown in Figure \ref{fig:fg2} (left) is 
the above cumulative employment rate and 
it increases monotonically up to $(1-U)_{n}=1$ as the stage $n$ goes on.  
On the other hand, 
the behavior of 
stage-wise unemployment rate $U^{(n)}$ might be more complicated 
and it contains useful information about 
the difficulties of 
matching between students and companies in each job-hunting process. 
However, unfortunately, 
there is no empirical data conferring the stage-wise unemployment rate 
opened to the public and we should estimate the quantities 
from the available data such as $(1-U)_{n}$. 
\subsection{Global mismatch and its measurement}
We should notice that 
in our model system, one has three relevant quantities, 
namely, 
unemployment rate $U^{(n)}$, 
labor shortage ratio $\Omega^{(n)}$ and 
job offer ratio $\alpha^{(n)}$. 
To obtain a useful relationship between these quantities, 
let us define the total number of students and companies 
by $N$ and $K$, 
respectively. Each student and company are 
specified by labels $i=1,\cdots, N; 
k=1,\cdots, K$, 
and let $v_{k}^{*}$ be the quota of the company 
$k$ and we define 
the number of students whom the company $k$ obtains by 
$m_{k}$. Then, we have 
\begin{equation}
\Omega= \frac{1}{V} \sum_{k=1}^{K}(v_{k}^{*}-m_{k})=1-\frac{1}{V}\sum_{k=1}^{K}v_{k}^{*}=1-\alpha^{-1}(1-U),
\end{equation}
where we used the definitions of 
total vacancy 
\begin{equation}
V \equiv \sum_{k=1}^{K}
v_{k}^{*}
\end{equation}
and the job offer ratio 
\begin{equation}
\alpha \equiv V/N.
\end{equation}
Apparently, for $\alpha  =V/N >1$, the labor market behaves as a `seller's market', whereas for $\alpha < 1$, the market becomes a `buyer's market'.
Therefore, 
one confirms that 
$U, \Omega$ and $\alpha$ should 
satisfy the following single equation \cite{Chen,Chenb}. 
\begin{equation}
U = 
\alpha \Omega + 
1-\alpha
\label{eq:U_alpha}
\end{equation}
By solving the above equation 
with respect to $\Omega$, we have the alternative form
\begin{equation}
\Omega = 
\frac{\alpha -(1-U)}{\alpha}. 
\label{eq:OU}
\end{equation}
By using this equation 
(\ref{eq:OU}), 
we calculate the labor shortage ratio $\Omega$ 
by means of the observables 
$1-U$ and $\alpha$ as we showed in Figure \ref{fig:fg2}. 
Here it should be kept in mind 
that the above relation is satisfied for 
cumulative 
quantities although we did not mention it clearly. 
Namely, (\ref{eq:OU}) (or equation (\ref{eq:U_alpha})) should be written as 
\begin{equation}
\Omega_{n} = \frac{\alpha- (1-U)_{n}}{\alpha}
\label{eq:Omega_n}
\end{equation}
where $\Omega_{n}$ stands for 
the cumulative labor shortage ratio. 
It could be defined explicitly as 
\begin{equation}
\Omega_{n} = 
\frac{
\mbox{
$\#$ of positions which are opened for students until the stage $n$}
}
{
\mbox{
total $\#$ of positions in society}
},
\end{equation}
which is different from the 
stage-wise labor shortage ratio 
\begin{equation}
\Omega^{(n)} = 
\frac{
\mbox{
$\#$ of positions which are still opened for students even after $n$-stage is over}
}
{
\mbox{
$\#$ of positions which are opened for students 
after $(n-1)$-stage is over}
}.
\end{equation}
\begin{figure}[!htb]
\resizebox{\textwidth}{!}{%
  \includegraphics{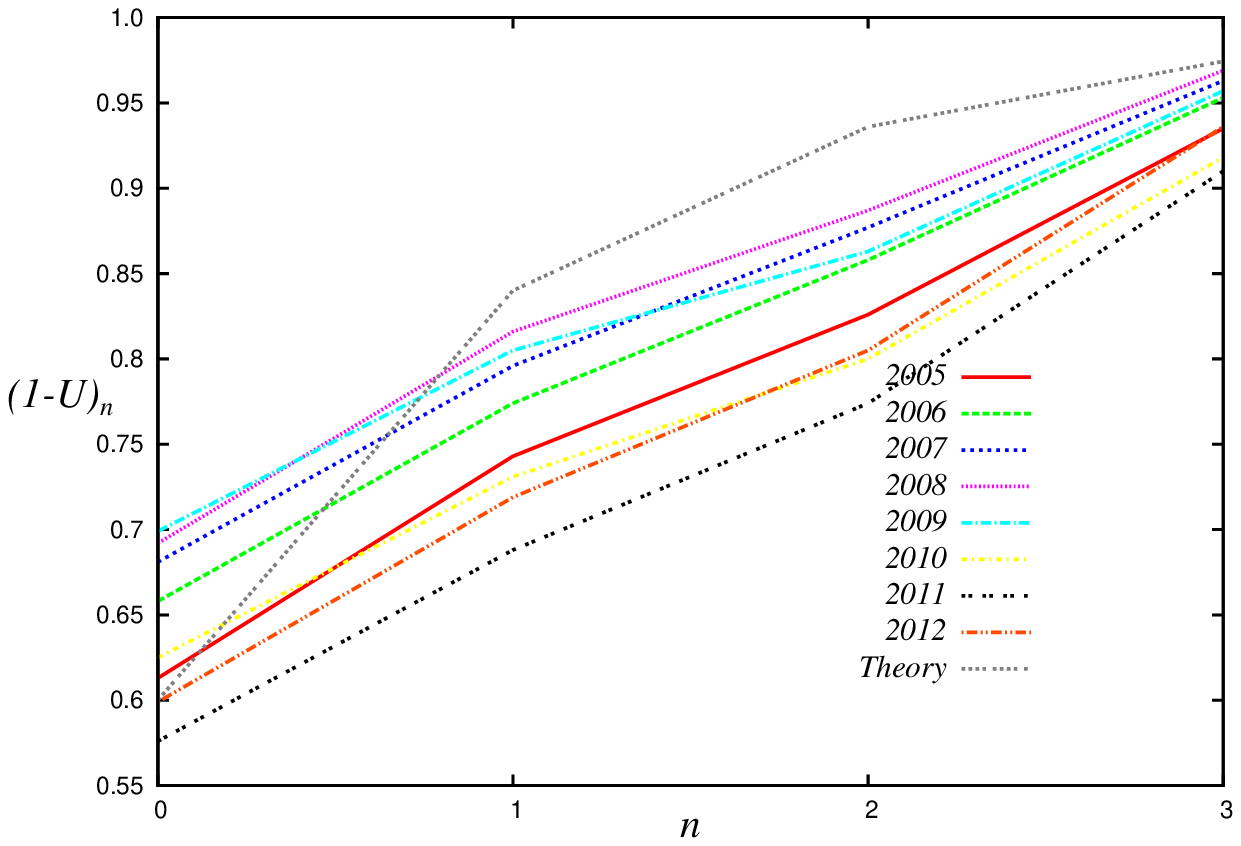}
 \includegraphics{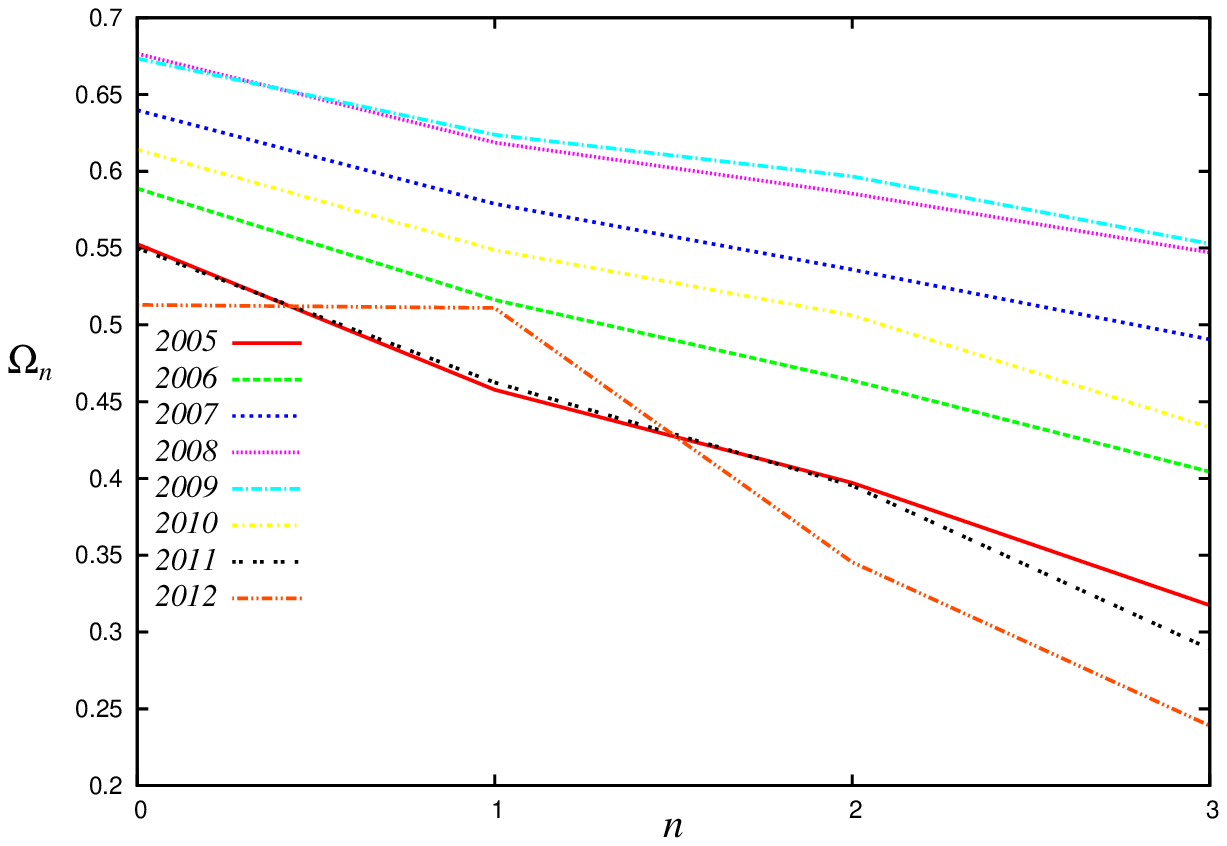}
}
\caption{\footnotesize 
The cumulative employment rate $(1-U)_{n}$ (left) and 
labor shortage ratio $\Omega_{n}$ (right) evaluated by 
employment statistical data of university graduates in 1996 to 2012 by MEXT \& MHLW \cite{mext}. 
$\Omega_{n}$ is calculated by $\Omega_{n}=\{\alpha-(1-U)_{n}\}/\alpha$. 
A line caption `Theory' in the left panel denotes $(1-U)_{n}=1-U^{n}$ 
which is obtained from the assumption $U^{(n)}=U=0.4$. 
From this panel, we find that 
the empirical data does not follow 
the theoretical law: $1-U^{n}$. 
}
\label{fig:fg22}
\end{figure}
In Figure \ref{fig:fg22}, we plot the cumulative employment rate $(1-U)_{n}$ (left) and 
labor shortage ratio $\Omega_{n}$ (right) evaluated by 
employment statistical data of university graduates in 1996 to 2012 by MEXT \& MHLW \cite{mext}. 
From the right panel, we are confirmed that 
the $(1-U)_{n}$ increases monotonically as the stage $n$ gains. 
On the other hand, the cumulative labor shortage ratio $\Omega_{n}$ 
is apparently a monotonically increasing function of $n$.  

In following, we write 
$\Omega_{n} \equiv \Omega, (1-U)_{n} \equiv 1-U$ 
for simplicity and 
draw the $U$-$\Omega$ curve using 
the empirical data for the cumulative employment rate $1-U$.

\begin{figure}[!htb]
\resizebox{\textwidth}{!}{%
 \includegraphics{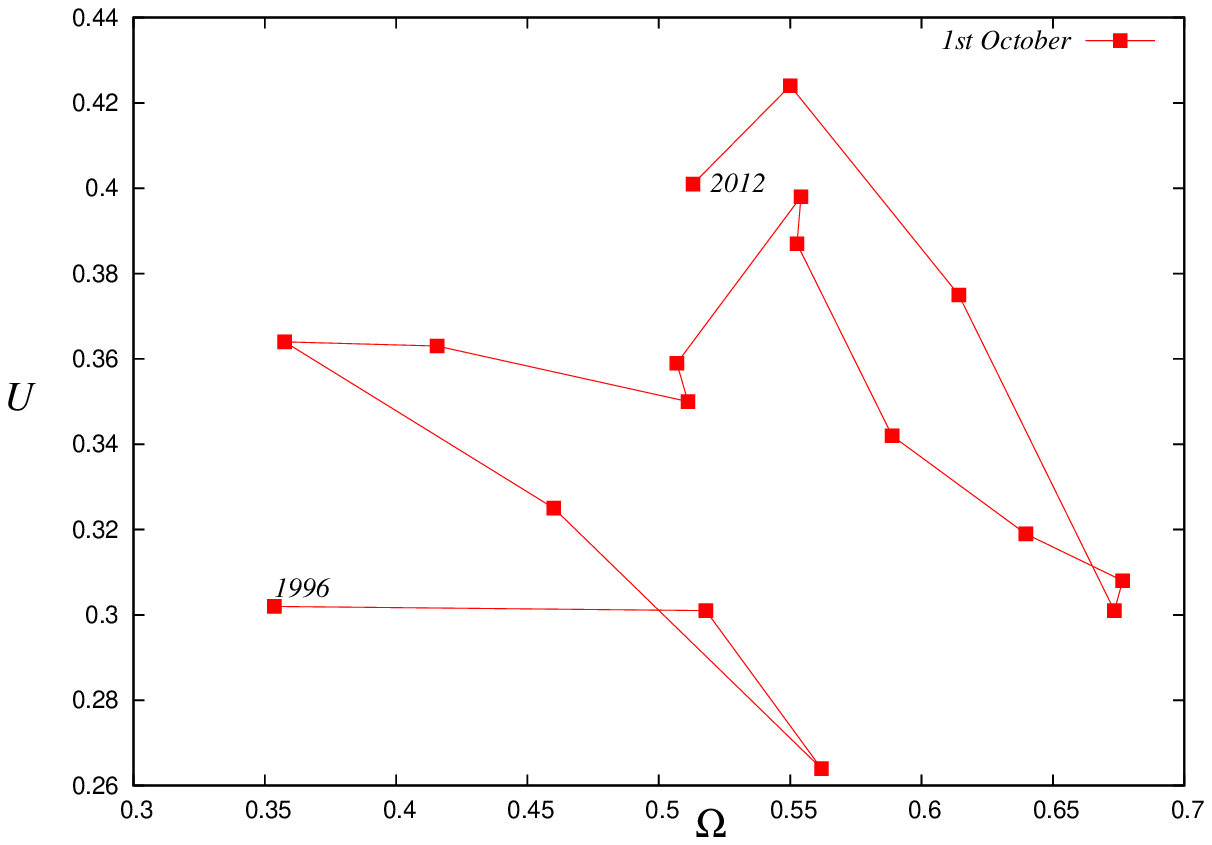}
  \includegraphics{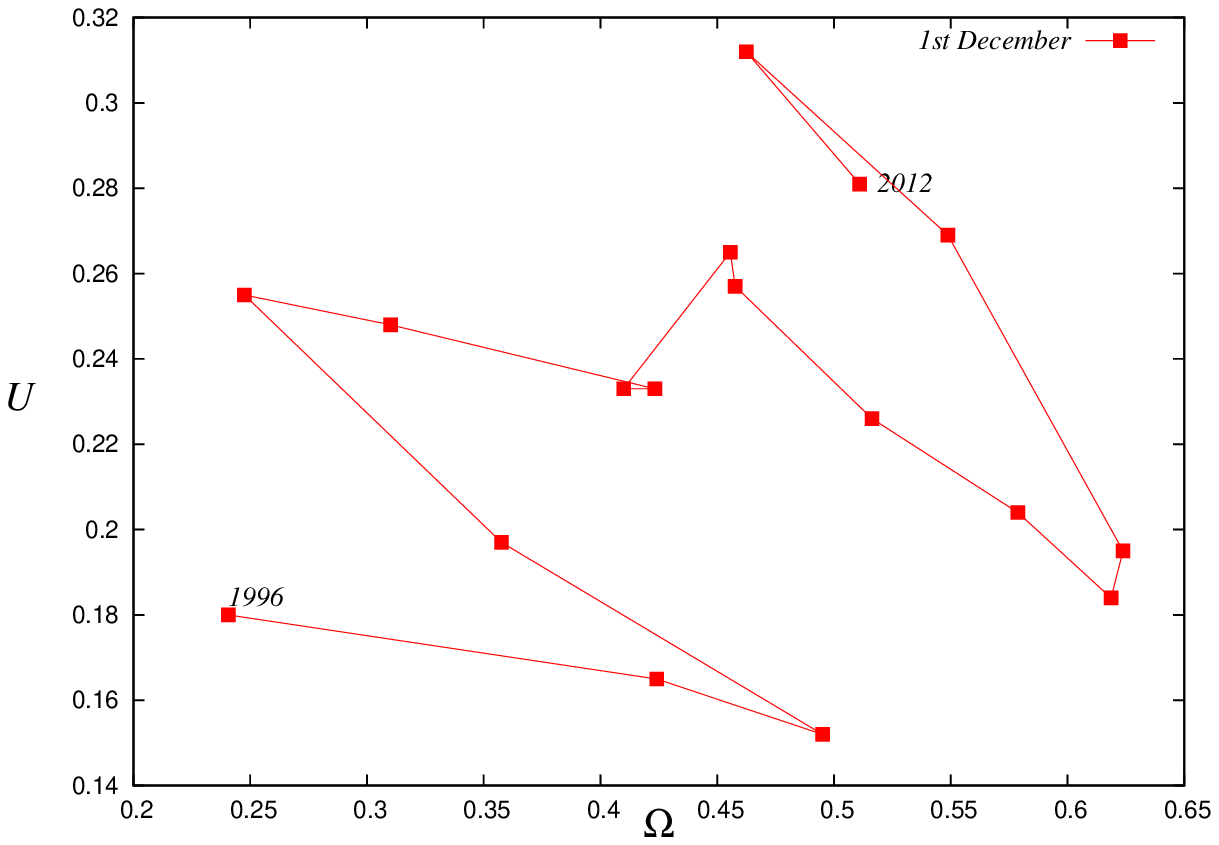}
} \\
\resizebox{\textwidth}{!}{%
 \includegraphics{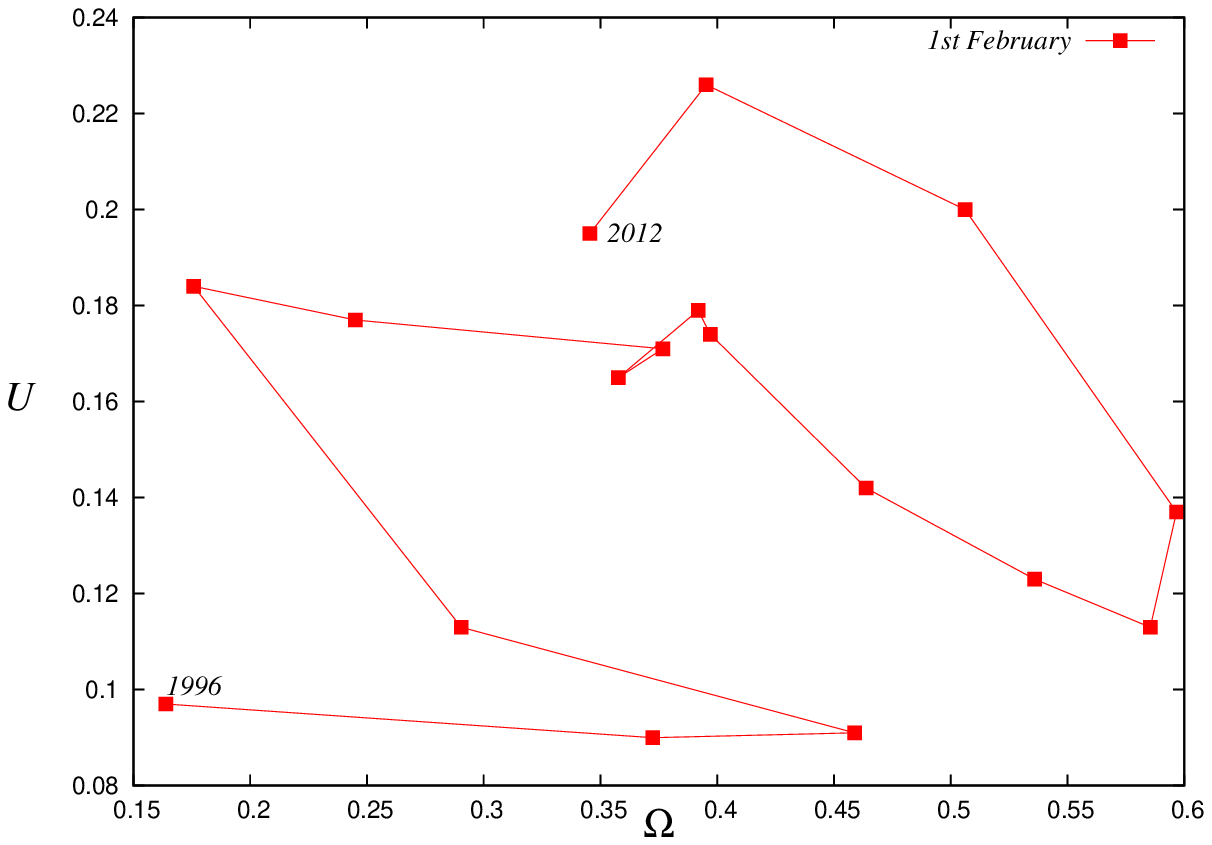}
  \includegraphics{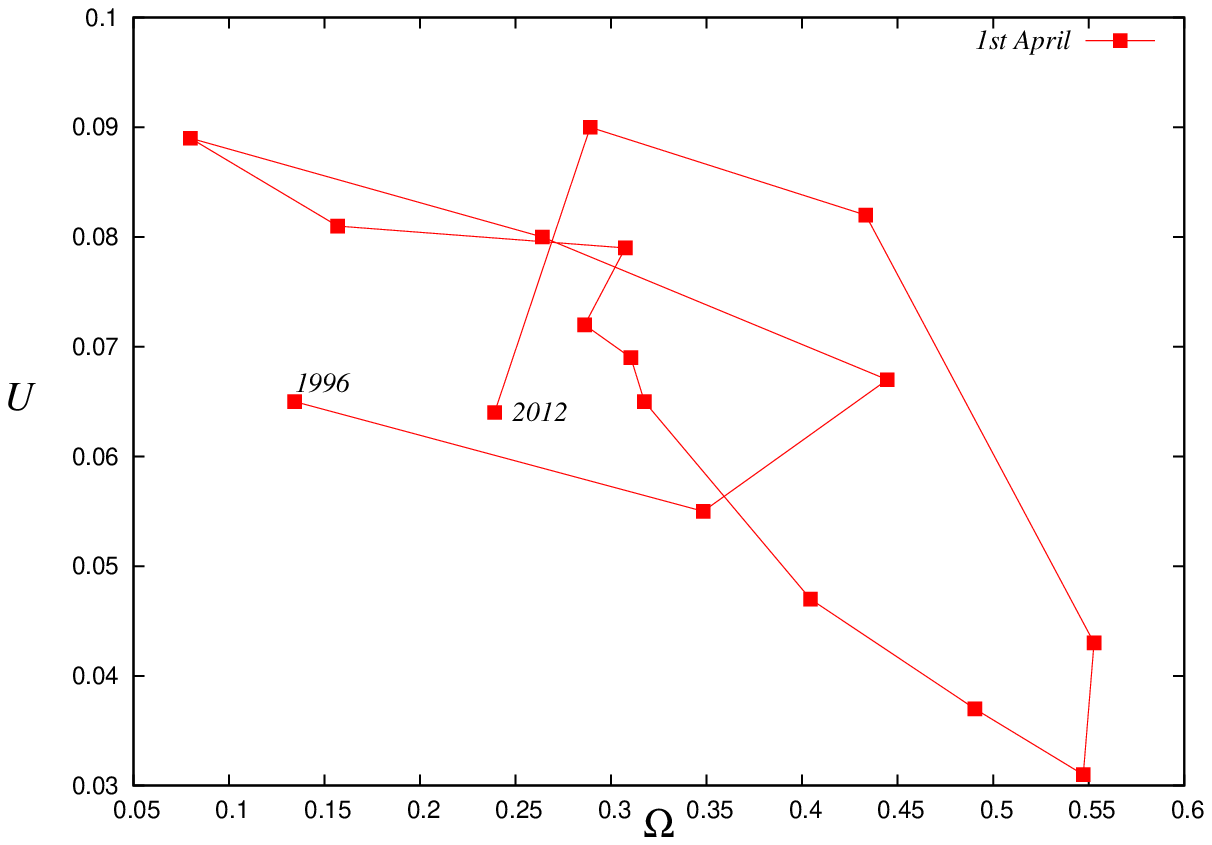}
}
\caption{\footnotesize 
The $U$-$\Omega$ curves for 
October, December, February and April, 
which are calculated by empirical data of 
$1-U$ \cite{mext} 
and equation (\ref{eq:OU}). 
We find that the degree of global mismatch 
between companies and students 
in 2010 is the worst in last seventeen years. 
However, 
in 2011 and 2012, 
the mismatch is even improved by compromise 
between companies and students. 
}
\label{fig:fg001}
\end{figure}

In Figure \ref{fig:fg001}, we show 
the $U$-$\Omega$ curves for 
seventeen years in 
October, December, February and April, 
which are calculated by using the data \cite{mext}.  
From these four panels, 
we find that the degree of global mismatch 
between companies and students 
in 2010 is the worst in the past seventeen years. 
However, 
in 2011 and 2012, 
the mismatch is even improved by compromise 
between companies and students. 

As we have shown above, 
the $U$-$\Omega$ curve 
tells us useful information about the global mismatch. 
Especially, the trajectory as shown in Figure \ref{fig:fg001} 
is reflected by changing students' tendencies for the choice of companies, 
government's policy, or business cycle in the society.  
\subsection{Representation of $(\alpha^{(n)},U^{(n)},\Omega^{(n)})$ in terms of $(1-U)_{n}, (1-U)_{n-1}$ and $\alpha$}
In the previous section, we showed 
the $U$-$\Omega$ curve 
for 
the cumulative quantities 
$(1-U)_{n}$ and $\Omega_{n}$. 
From the result, we were confirmed that 
global mismatch between students and companies was actually enhanced 
around 2010. 
However, 
the information about the 
$U$-$\Omega$ curve 
does not tell us 
to what extent the market becomes 
difficult or easy for students to obtain the jobs, 
or to what extent the market is sufficient or insufficient for 
companies to fill their quota as the stage $n$ goes on. 
In order to investigate this kind of 
macroscopic properties of the market, 
we should draw 
the $U$-$\Omega$ curve for 
`stage-wise' quantities 
instead of the cumulative ones. 

Then, we should notice that 
the relationship (\ref{eq:OU}) (or equation (\ref{eq:U_alpha})) should be satisfied 
for the `stage-wise' quantities $(\alpha^{(n)},U^{(n)},\Omega^{(n)})$ as well. 
Hence, by setting 
$\Omega =\Omega^{(n)}, 
U =U^{(n)}$ and $\alpha=\alpha^{(n)}$, 
we obtain a single relationship: 
\begin{equation}
U^{(n)}=\alpha^{(n)} \Omega^{(n)} + 1-\alpha^{(n)}.
\label{eq:UOmegaalpha}
\end{equation}
We should keep in mind that 
$\alpha^{(0)}=\alpha, 
U^{(0)}=U_{0}=U, 
\Omega^{(0)}=\Omega_{0}=\Omega$, 
and the stage-wise job-offer ratio $\alpha^{(n)}$ is defined as 
 \begin{equation}
 \alpha^{(n)} = 
 \frac{
 \mbox{
$\#$ of positions opened for students after $n$-stage is over}
}
{
\mbox{
$\#$ of students who are looking for the position after $n$-stage is over}
}. 
\end{equation}
From the definition, $\alpha \equiv \alpha^{(0)}$ appearing 
in equation (\ref{eq:OU}) for the cumulative quantities should be 
$\alpha^{(0)}=V/N$ and 
the counter empirical data is opened to the public in the October's statistics. 
 
Here we attempt to 
rewrite the set of relevant quantities 
$(\alpha^{(n)},U^{(n)},\Omega^{(n)})$ in the stage-wise sense 
at the $n$-th job-hunting stage 
in terms of 
the cumulative 
employment rates $(1-U)_{n}, (1-U)_{n-1}$ and job-offer ratio $\alpha \equiv \alpha^{(0)}$. 
For this purpose, 
we first examine the expression for $\alpha^{(1)}$. 

The total positions (vacancies) after the first stage is over, 
namely, $V^{(1)}$ is given by 
\begin{equation}
V^{(1)} = 
V^{(0)} -(1-U^{(0)}) N  =  \{
\alpha -1 + U^{(0)}
\}N
\end{equation}
where we used 
$V \equiv V^{(0)}=\alpha N$. 
On the other hand, 
the number of students who could not get any position at $n=1$  is given 
as $N^{(1)} \equiv NU^{(0)}$. Hence, we have 
\begin{equation}
\alpha^{(1)} = 
\frac{V^{(1)}}{N^{(1)}}=
\frac{\{\alpha - 1+ U^{(0)}\}N}
{U^{(0)}N} = 
\frac{\alpha -1 + U^{(0)}}{U^{(0)}}. 
\end{equation}
Using the same way as the above $V^{(1)}$, 
we obtain the number of positions still remaining after the second stage as 
\begin{eqnarray}
V^{(2)} & = & 
V^{(1)} - (1-U^{(1)})U^{(0)}N \nonumber \\
\mbox{} & = & 
V-(1-U^{(0)})N-(1-U^{(1)})U^{(0)}N = 
\{
\alpha -1 + U^{(1)}U^{(0)}
\}N. 
\end{eqnarray}
Therefore, when we notice that 
$N^{(2)}=U^{(1)}U^{(0)}N$, 
one can rewrite the  
$\alpha^{(2)}$ in terms of $U^{(0)},U^{(1)}$ and $\alpha$ as follows. 
\begin{equation}
\alpha^{(2)} = 
\frac{V^{(2)}}
{N^{(2)}} = 
\frac{\alpha -1 + U^{(1)}U^{(0)}}
{U^{(1)}U^{(0)}} 
\end{equation}
Repeating this procedure up to 
$n$-stage, we easily obtain 
\begin{equation}
\frac{V^{(n)}}{N} = 
\alpha -1 + 
\prod_{k=0}^{n-1}
U^{(k)},\,\, \frac{N^{(n)}}{N}=\prod_{k=0}^{n-1}U^{(k)}
\end{equation}
and this simply reads 
\begin{equation}
\alpha^{(n)} = 
\frac{V^{(n)}}{N^{(n)}} = 
\frac{\alpha - 1 + \prod_{k=0}^{n-1}U^{(k)}}
{\prod_{k=0}^{n-1}U^{(k)}}\,\, (n \geq 1),\,\,\, \alpha^{(0)}=\alpha. 
\label{eq:alpha_n1}
\end{equation}

We next rewrite 
the above $\alpha^{(n)}$ by means of 
the cumulative employment rate  
$(1-U)_{n-1}$ which is available 
as an empirical data set shown in Figure \ref{fig:fg2}. 
Obviously, 
$V^{(n)}$ is given in terms of $(1-U)_{n-1}$ as 
$V^{(n)} = 
\{
\alpha -(1-U)_{n-1}
\}N$. 
On the other hand, the number of 
students who could not get any job until $n$-stage is 
evaluated by $N^{(n)}=N-(1-U)_{n-1}N=\{1-(1-U)_{n-1}\}N$. 
Thus, we have 
\begin{equation}
\alpha^{(n)} = 
\frac{V^{(n)}}{N^{(n)}} = 
\frac{\alpha -(1-U)_{n-1}}{1-(1-U)_{n-1}}. 
\label{eq:alpha_n2}
\end{equation}
By comparing equations (\ref{eq:alpha_n1}) and (\ref{eq:alpha_n2}), we immediately obtain
\begin{equation}
(1-U)_{n-1}=1-\prod_{k=0}^{n-1}U^{(k)}. 
\label{eq:1-U_n}
\end{equation}
In order to derive the expression of `stage-wise' unemployment rate $U^{(k)}$ in terms of 
the cumulative employment rate $(1-U)_{k}$, we shall write down (\ref{eq:1-U_n}) for $n=2,1$ and $n=0$. 
This procedure immediately leads to 
$(1-U)_{2} = 
1-U^{(0)}U^{(1)}U^{(2)}, 
(1-U)_{1} = 
1-U^{(0)}U^{(1)}$ and 
$(1-U)_{0} = 1-U^{(0)}$. 
Solving these equations with respect to 
$U^{(k)},k=0,1,2$, one obtains 
\begin{equation}
U^{(0)} = 
1-(1-U)_{0},\,\,
U^{(1)} = 
\frac{1-(1-U)_{1}}{1-(1-U)_{0}},\,\,
U^{(2)} = 
\frac{1-(1-U)_{2}}{1-(1-U)_{1}}. 
\end{equation}
The above result is easily generalized to $U^{(n)}$ as 
\begin{equation}
U^{(n)} = 
\frac{1-(1-U)_{n}}{1-(1-U)_{n-1}}. 
\end{equation}
It is quite easy for us to 
prove that $(0 \leq )\, U^{(n)} \leq 1$ holds. 
When we notice that 
the cumulative employment is 
a monotonically increasing 
function of $n$, one has 
$(1-U)_{n-1} \leq (1-U)_{n}$,  
that is, 
$1-(1-U)_{n} \leq 1-(1-U)_{n-1}$. 
Hence, we conclude 
\begin{equation}
(0 \leq )\, \frac{1-(1-U)_{n}}{1-(1-U)_{n-1}} \equiv U^{(n)} \leq 1. 
\label{eq:U_bound}
\end{equation}
Using the relation (\ref{eq:UOmegaalpha}), we immediately obtain 
\begin{equation}
\Omega^{(n)}  =  
\frac{U^{(n)}+\alpha^{(n)}-1}{\alpha^{(n)}} =
\frac{\alpha - (1-U)_{n}}{\alpha -(1-U)_{n-1}}.
\end{equation}
From the fact that 
$(1-U)_{n}$ decreases monotonically as $n$ increases, 
we can easily show $(0 \leq)\,\,\Omega^{(n)} \leq 1$ using the same argument 
as in $U^{(n)}$ (the above equation (\ref{eq:U_bound})). 

Let us summarize the result. 
\begin{equation}
(\alpha^{(n)},U^{(n)},\Omega^{(n)}) = 
\left(
\frac{\alpha-(1-U)_{n-1}}
{1-(1-U)_{n-1}}, 
\frac{1-(1-U)_{n}}
{1-(1-U)_{n-1}},
\frac{\alpha - (1-U)_{n}}
{\alpha - (1-U)_{n-1}}
\right)
\label{eq:three}
\end{equation}
In following, 
we shall investigate the generic properties 
of these stage-wise quantities. 
\subsubsection{On the monotonic behavior of stage-wise job-offer ratio}
\begin{figure}[!htb]
\resizebox{\textwidth}{!}{%
  \includegraphics{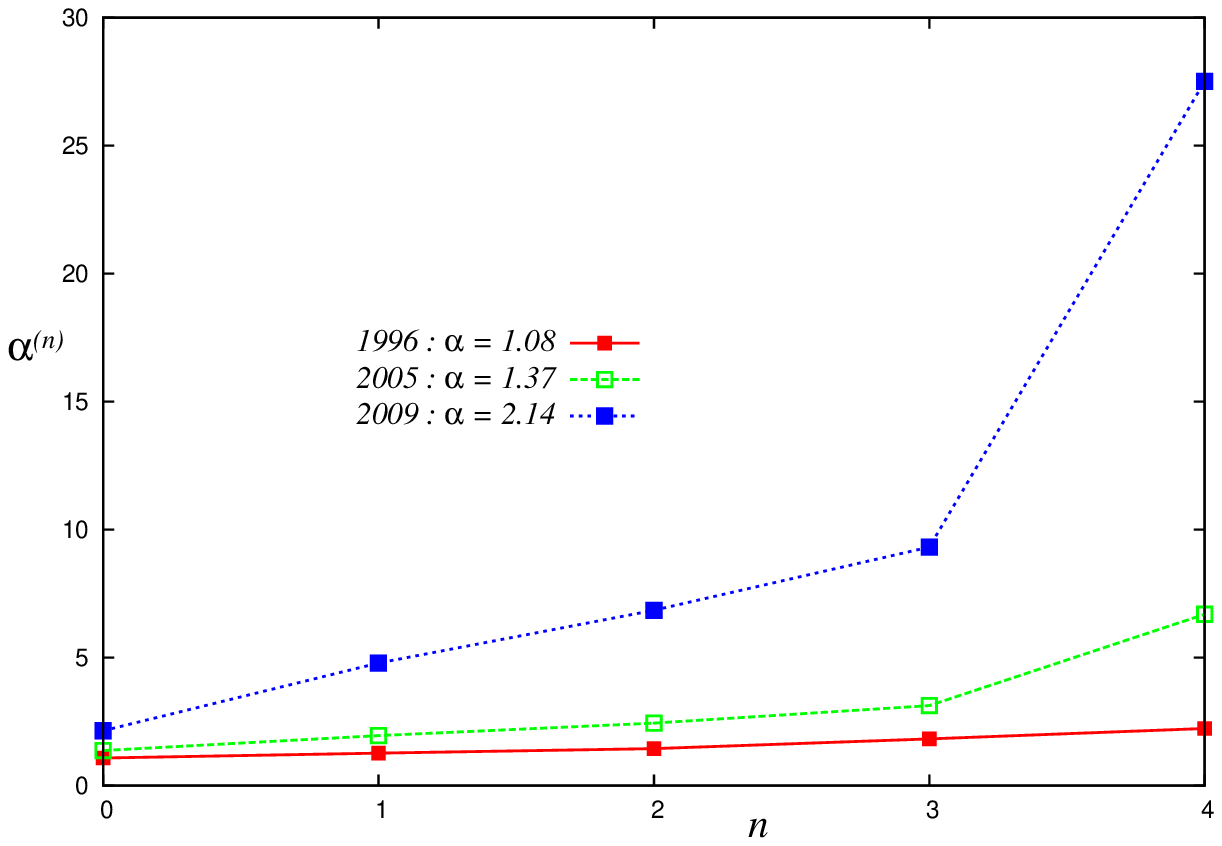}
 \includegraphics{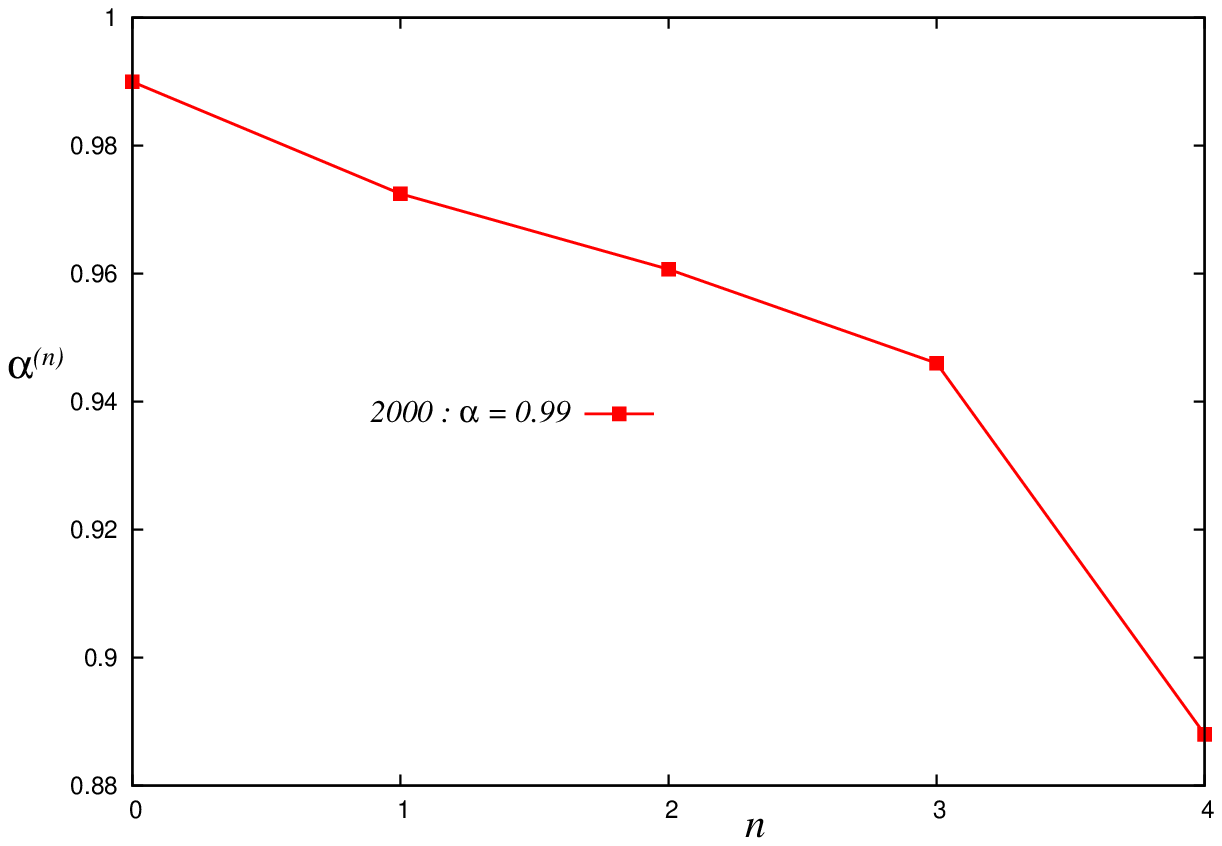}
}
\caption{\footnotesize 
The stage-wise job-offer ratio $\alpha^{(n)}=\{\alpha - (1-U)_{n-1}\}/\{1-(1-U)_{n-1}\}$. 
The right panel shows several typical examples for the case of 
$\alpha >1$ (for the statistics in 1996, 2005 and 2009), whereas the right panel is shown as a result for $\alpha <1$ 
(for the statistics in 2000). 
}
\label{fig:fg222B}
\end{figure}

It should bear in mind that 
the behavior of the stage-wise job-offer ratio $\alpha^{(n)}$ is strongly dependent on 
the initial value $\alpha^{(0)}=\alpha$. 
Actually, one can evaluate 
the difference (gap) $\alpha^{(n)}-\alpha^{(n-1)}$ as 
\begin{equation}
\alpha^{(n)}-\alpha^{(n-1)} =  
\frac{(\alpha-1)\{(1-U)_{n-1}-(1-U)_{n-2}\}}
{
\{
1-(1-U)_{n-1}
\}
\{
1-(1-U)_{n-2}
\}
}. 
\end{equation} 
Taking into account 
the the fact that 
the cumulative employment rate $(1-U)_{n}$ 
is a monotonically increasing function of stage $n$, 
namely, 
$(1-U)_{n-1} \geq (1-U)_{n-2}$, and it 
is lower than unity, 
that is, 
$(1-U)_{n-1}, (1-U)_{n-2} \leq 1$, 
we conclude that 
\begin{eqnarray*}
\alpha^{(n)} & > & \alpha^{(n-1)}\,\,\,\,\, (\alpha >1: \mbox{monotonically increasing}) \\
\alpha^{(n)} & < & \alpha^{(n-1)}\,\,\,\,\,(\alpha <1: \mbox{monotonically decreasing})
\end{eqnarray*}
Apparently, 
the stage-wise job-offer ratio is 
one of the measurement to quantify the 
difficulties of 
matching process between 
students and companies 
on the condition that one does not consider the 
extreme `biased-behavior' of the students' preferences. 
Therefore, 
the above result tells us that 
the job-hunting becomes more difficult as stage $n$ goes on 
when the job-offer ratio released to the public in October 
$\alpha^{(0)}=\alpha$ is lower than $1$, whereas 
it becomes easier as the stage $n$ increases when $\alpha >1$.   
Actually in Figure \ref{fig:fg222B}, we plot the $\alpha^{(n)}$ by making use of 
the empirical data. 
The left panel in Figure \ref{fig:fg222B} shows the results for 1996, 2005 and 2009 having $\alpha >1$, 
whereas the right panel is the result for the market in $2000$ having $\alpha=0.99 <1$. 
From this figure, we clearly find that 
the behavior of $\alpha^{(0)}$ is categorized into distinct two classes 
by $\alpha >1$ or $\alpha <1$. 
It should be stressed that 
this result is independent on 
the choice of concrete matching algorithm for 
collective intelligence of students in job-hunting process. 
We should mention that no similar monotonic behavior is observed in $\Omega^{(n)}$ or 
$U^{(n)}$. 

From the above general result, 
we obtain several insights into the behavior of recent Japanese labor market for 
university graduates. 
Very recently, in April 2013,  
a Japanese think tank, {\it Recruit Works Institute} \cite{Recruit} reported 
that the job-offer ratio in 2012 for the market at the first stage which 
includes all under-, graduate students  and all companies 
was $\alpha^{(0)} \equiv \alpha=1.28\, (>1)$. 
Hence, from the above discussion,
the stage-wise job-offer ratio $\alpha^{(n)}$ increases as $n$ increases. 
Thus, the students should not rush themselves to 
get the positions because 
it becomes much easier for them to obtain 
the jobs as $n$ goes on. 
However, 
the think tank also reported 
that the `effective' job-offer ratio was $\alpha=0.54\,(<1)$
when we restrict ourselves to 
the market of established and large-scale companies 
having more than $5000$ employees. 
This fact implies that 
the students who are looking for the positions in 
such established companies 
should hurry to obtain the jobs at as early-stage as possible 
because  
the stage-wise $\alpha^{(n)}$ decreases monotonically to zero for $\alpha <1$. 
On the other hand, 
the effective 
job-offer ratio for the market in which 
there are only small enterprises having less than 
$300$ employees was $3.26$ in 2012. 
This fact might be a reason why a global mismatch 
is difficult to be overcame in recent Japanese labor markets 
for university graduates. 
\subsubsection{The asymptotic limit}
As we already explained in the introduction, 
in Japan, 
the statistics for the cumulative employment rate is 
opened to the public in four seasons, 
namely, 
in October ($n=0$), 
December ($n=1$), 
February ($n=2$) and 
April ($n=3$). 
However, 
it is possible for us to imagine 
the case of quite large $n$, especially, 
the asymptotic limit of 
stage-wise quantities 
$(\alpha^{(n)}, U^{(n)}, \Omega^{(n)})$ in 
the limit of $n \to \infty$. 
Hence, here we shall consider the limit $n \to \infty$ in the results (\ref{eq:three}) 
obtained in the previous sections.
 
We first consider the case of $\alpha <1$. 
To consider the limit $n \to \infty$, we should notice that 
the number of students who get their jobs becomes identical to 
the total vacancy $V$ in the limit of $n \to \infty$ for $\alpha <1$. 
Namely
\begin{equation}
\lim_{n \to \infty}(1-U)_{n}=\frac{V}{N}=\alpha. 
\label{eq:alpha_leq}
\end{equation}
In this sense, the cumulative employment rate 
could not increases up to unity and it 
is apparently bounded from the above by 
the job-offer ratio opened in October ($n=0$), 
namely, $\alpha^{(0)}=\alpha$. 
Therefore, we easily confirm at the end 
\begin{equation}
\lim_{n \to \infty}\alpha^{(n)} =0, \,\,
\lim_{n \to \infty}U^{(n)}=1, \,\,
\lim_{n \to \infty}\Omega^{(n)}=0
\end{equation}
where we used the fact 
\begin{equation}
\lim_{n \to \infty}
\{(1-U)_{n}-(1-U)_{n-1}\}=\epsilon
\label{eq:infinitesimal}
\end{equation}
for 
an infinitesimal constant $\epsilon(>0)$ accompanying with 
equation (\ref{eq:alpha_leq}) to derive the above second and third limits as 
\begin{eqnarray}
\lim_{n \to \infty}U^{(n)} & = &  
\lim_{n \to \infty}
\frac{1-(1-U)_{n}}
{1-(1-U)_{n-1}}  =  
\lim_{n \to \infty}
\frac{1-(1-U)_{n}}{1-(1-U)_{n} + \epsilon}=
\frac{1-\alpha}{1-\alpha + \epsilon}=1, \\
\lim_{n \to \infty}
\Omega^{(n)} & = & 
\lim_{n \to \infty}
\frac{\alpha - (1-U)_{n}}
{\alpha - (1-U)_{n-1}}  = 
\lim_{n \to \infty}
\frac{\alpha -(1-U)_{n}}
{\alpha -(1-U)_{n}+ \epsilon}= \frac{0}{\epsilon}=0. 
\end{eqnarray}

We next consider the case of $\alpha > 1$. 
For this case, 
all students can get the job 
in the limit of $n \to \infty$. 
Hence, we have 
\begin{equation}
\lim_{n \to \infty}
(1-U)_{n} = 1. 
\label{eq:1-U_n2}
\end{equation}
By taking into account this fact, we immediately obtain 
\begin{equation}
\lim_{n \to \infty} \alpha^{(n)}=\infty, \,\,
\lim_{n \to \infty}U^{(n)}=0,\,\,
\lim_{n \to \infty}\Omega^{(n)} =1
\end{equation}
where we used the relation 
(\ref{eq:infinitesimal}) and (\ref{eq:1-U_n2}) to get the second and third limits as 
\begin{eqnarray}
\lim_{n \to \infty}U^{(n)} & = & 
\lim_{n \to \infty}
\frac{1-(1-U)_{n}}{1-(1-U)_{n}+\epsilon}=\frac{0}{\epsilon}=0,\\
\lim_{n \to \infty}
\Omega^{(n)} & = & 
\lim_{n \to \infty}
\frac{\alpha -(1-U)_{n}}
{\alpha - (1-U)_{n}+ \epsilon} = 
\frac{\alpha -1}{\alpha -1 + \epsilon}=1. 
\end{eqnarray}
\subsubsection{The marginal $\alpha=1$ as a special case}
In the previous sections, 
we investigated the behavior of 
$\alpha^{(n)}$ for both cases of $\alpha >1$ and $\alpha<1$. 
Here we would like to stress that 
the marginal case $\alpha=1$ is rather special in the sense that 
$\alpha^{(n)}$ becomes independent of $n$ as 
\begin{equation}
\alpha^{(n)}=1. 
\end{equation}
This result is intuitively understood as follows. 
Let us define 
$\Delta_{V}^{(n)}$ and $\Delta_{N}^{(n)}$ as the number of 
positions which are occupied and the number of 
students who obtain the position at $n$-stage, respectively. 
Then, $\alpha^{(n)}$ is given as 
\begin{equation}
\alpha^{(n)} \equiv 
\frac{V^{(n)}}{N^{(n)}}=
\frac{N-\Delta_{V}^{(1)}-\Delta_{V}^{(2)}-\cdots-\Delta_{V}^{(n-1)}}
{N-\Delta_{N}^{(1)}-\Delta_{N}^{(2)}-\cdots-\Delta_{N}^{(n-1)}}=1
\end{equation}
where we used $V=\alpha N=N$.  
We also utilized the fact that the number of students who obtain the positions 
at each stage is identical to 
the number of positions occupied by those students, 
namely,  $\Delta_{V}^{(k)}=\Delta_{N}^{(k)},\,k=1,\cdots,n-1$. 

From the result (\ref{eq:three}), we also confirm 
\begin{equation}
U^{(n)}=\Omega^{(n)}
\end{equation}
for $\alpha=1$. 
Thus, 
the performance of 
students to find the positions is identical to 
that of companies to fill 
the quota at each stage $n$. 
\begin{figure}[!htb]
\resizebox{\textwidth}{!}{%
  \includegraphics{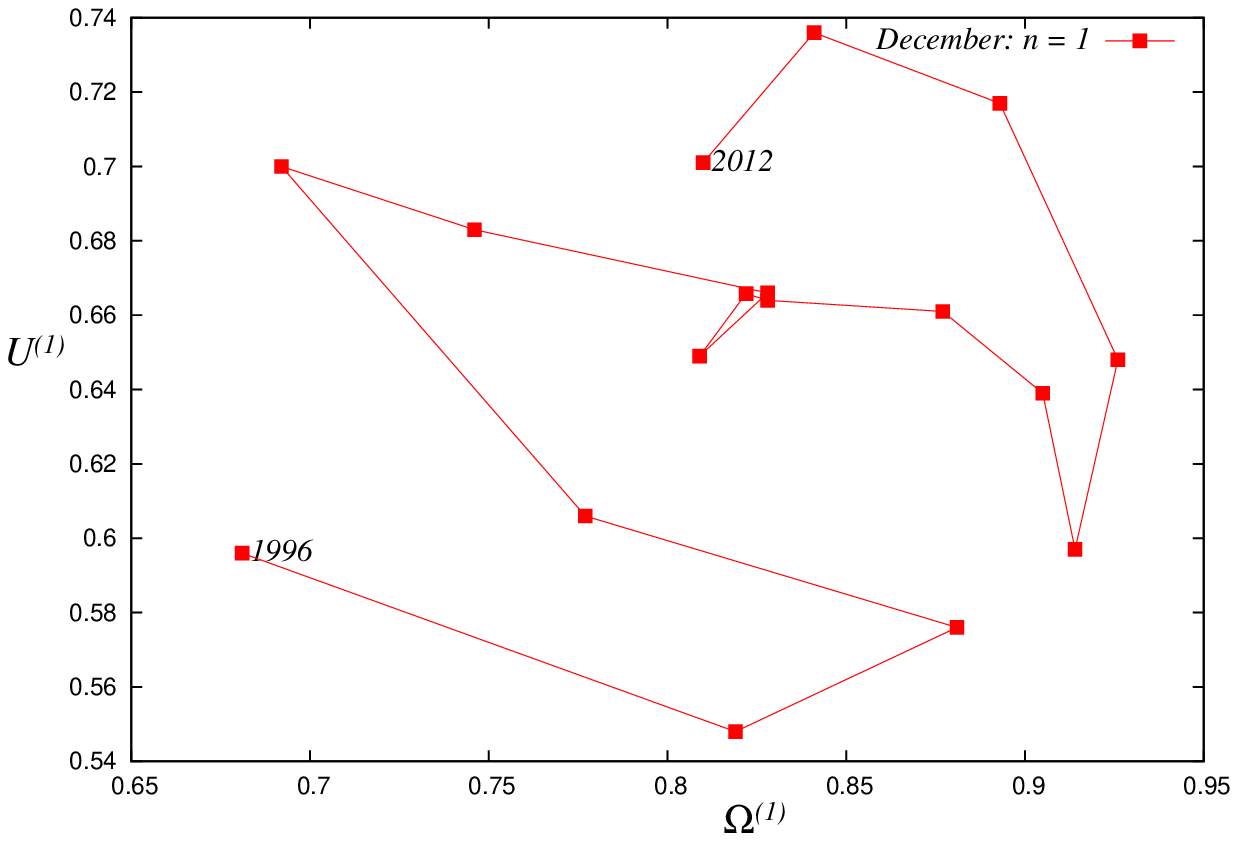}
 \includegraphics{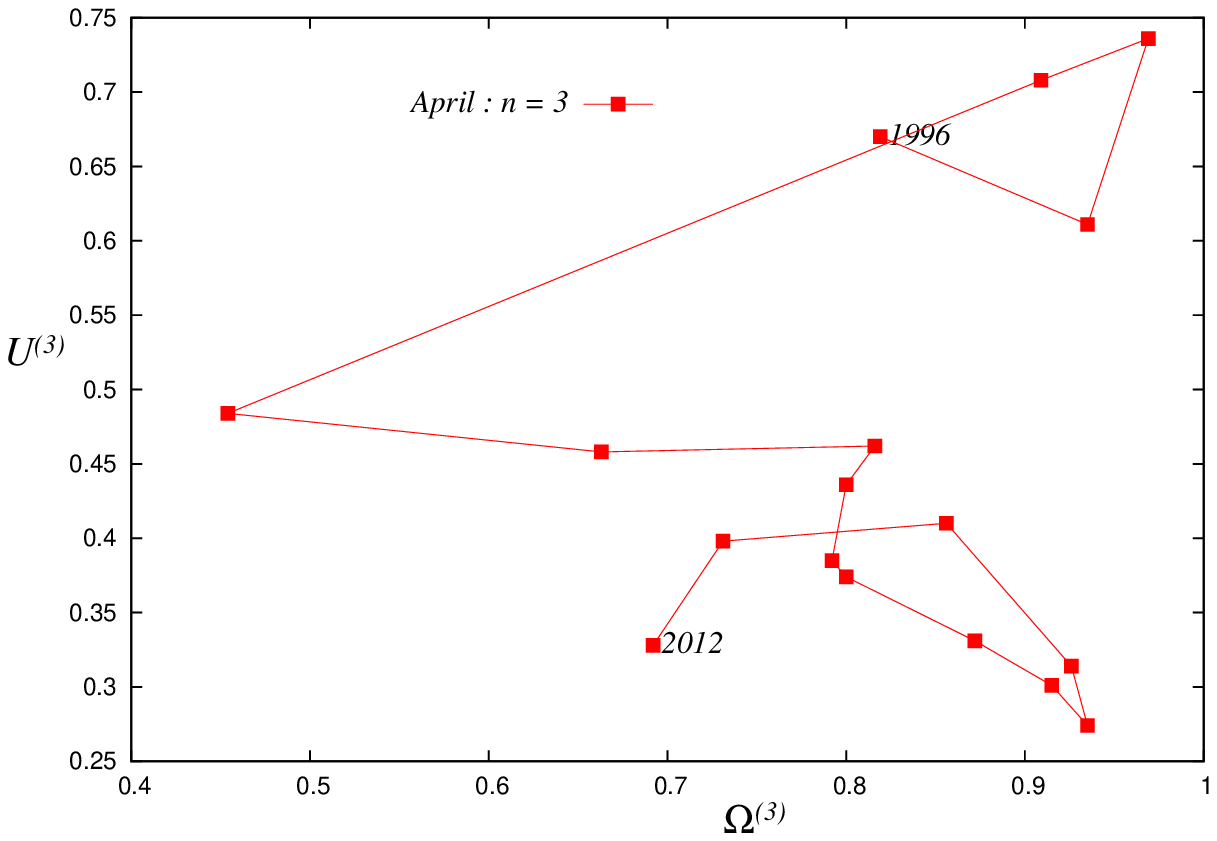}
}
\caption{\footnotesize 
The $U$-$\Omega$ curves evaluated 
by means of stage-wise quantities. 
Namely, $U^{(1)}$-$\Omega^{(1)}$ curve (left) and 
$U^{(3)}$-$\Omega^{(3)}$ curve (right). 
The trajectories $(\Omega_{\rm year}^{(n)},U_{\rm year}^{(n)}), {\rm year}=1996,1997, \cdots, 
2012$ are calculated in terms of the cumulative employment rate $(1-U)_{n}$ by equation (\ref{eq:three}). 
}
\label{fig:fg222}
\end{figure}
\mbox{}

In Figure \ref{fig:fg222}, 
we plot the $U$-$\Omega$ curves evaluated 
by means of stage-wise quantities. 
Namely, $U^{(1)}$-$\Omega^{(1)}$ curve (left) and 
$U^{(3)}$-$\Omega^{(3)}$ curve (right). 
From the right panel accompanying with the upper left panel in Figure \ref{fig:fg001}, 
which is the $U^{(0)}$-$\Omega^{(0)}$ curve from the definition, 
we clearly find that in 1996, the lowest global mismatch in $U^{(0)}$-$\Omega^{(0)}$ curve is getting worse 
as the stage $n$ gains up to $n=3$ as shown in Figure \ref{fig:fg222} (right). From the result, we are confirmed that 
the global mismatch could be changed as the stage $n$ goes on and 
it implies that students might change their strategies (`softening' their policies in some sense) during their job-hunting processes to get the positions. 
\subsection{`Learning curve' for the collective intelligence}
In the previous sections, 
we derived  
the stage-wise quantities 
$(\alpha^{(n)},U^{(n)},\Omega^{(n)})$ in terms of 
the cumulative employment rate $(1-U)_{n}, (1-U)_{n-1}$ and $\alpha$. 
We also discussed several generic properties concerning these stage-wise quantities. 
However, we need a concrete `learning algorithm' of 
collective intelligence which consists of job-seeking students. 
If these job-seeking students are `intelligent' enough to obtain their positions 
as quick as possible. In other words, 
if the students' strategies are effective enough to fill the total vacancies 
at as early stage as possible, 
the `learning algorithm' should be regarded as the best one. 
As the result, 
the quantity: 
\begin{equation}
\varepsilon_{n} \equiv 
\left\{
\begin{array}{cc}
1-(1-U)_{n} & (\alpha \geq 1) \\
\alpha - (1-U)_{n} & (\alpha <1)
\end{array}
\right.
\label{eq:LC}
\end{equation}
which is regarded as an error-measurement for 
the collective intelligence, 
converges to zero at the earliest stage of $n$. 
Therefore, 
the $n$-dependence of 
the error-measurement $\varepsilon_{n}$ 
should be referred to as `learning curve'  (see {\it e.g.} \cite{Engle,Bishop}), which is usually used in 
the literature of machine learning,  for 
the behavior of collective intelligence of students. 

Here we should mention that 
from equation (\ref{eq:Omega_n}), 
the cumulative labor shortage ratio 
$\Omega_{n}$ is a monotonically decreasing function of $n$ because 
$(1-U)_{n}$ increases monotonically. 
Therefore, 
the asymptotic limit is given by 
\begin{equation}
\lim_{n \to \infty}
\Omega_{n} = 
\left\{
\begin{array}{cc}
0 & (\alpha \leq 1) \\
\frac{\alpha-1}{\alpha} & (\alpha >1) 
\end{array}
\right.
\end{equation}

Obviously, 
we need to provide a specific learning algorithm to 
evaluate the learning curve. 
In following, we examine the learning curve for 
a simplest scenario, 
namely, 
the stage-wise 
unemployment rate $U^{(n)}$ or 
labor shortage ratio $\Omega^{(n)}$ is 
independent of $n$, namely, $U^{(n)}=U$ or 
$\Omega^{(n)}=\Omega$. 
\subsubsection{A result from scale-invariance in $U$ or $\Omega$}
\begin{figure}[!htb]
\resizebox{\textwidth}{!}{%
 \includegraphics{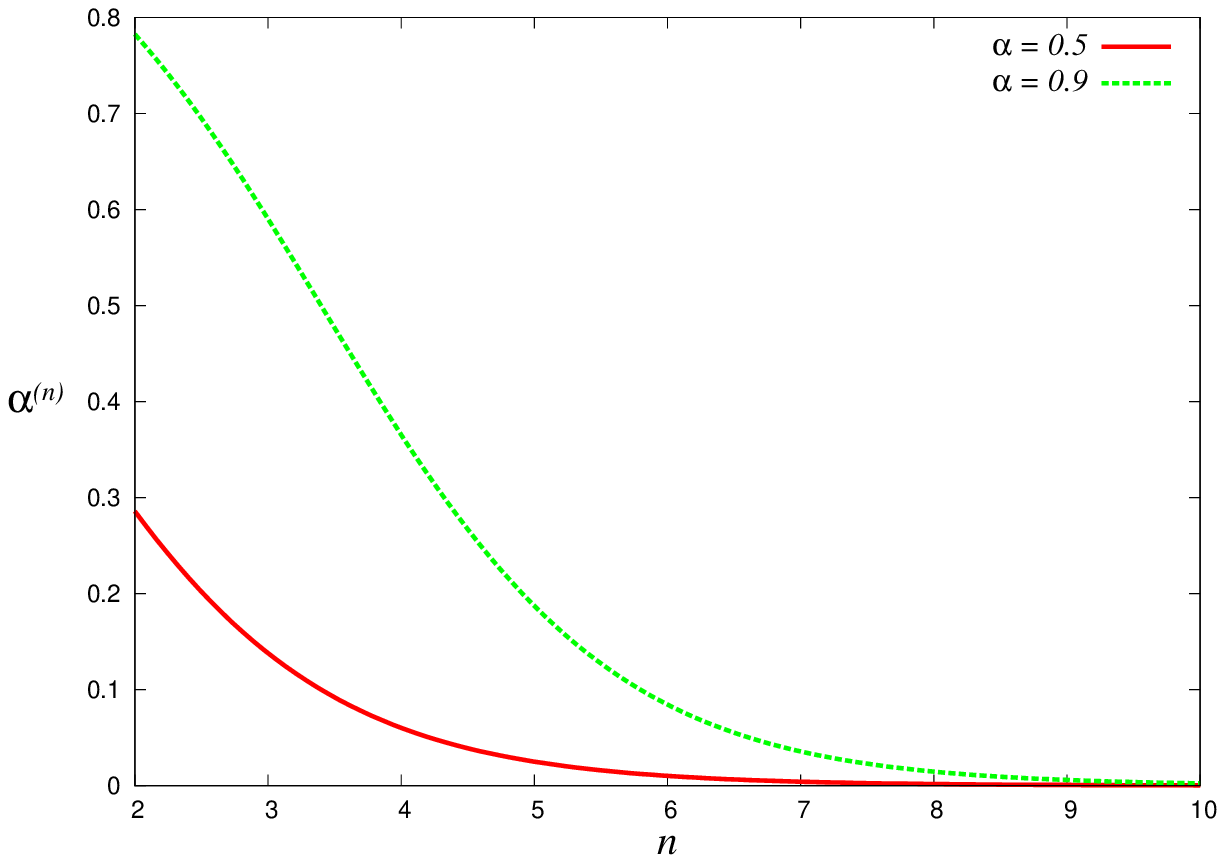}
  \includegraphics{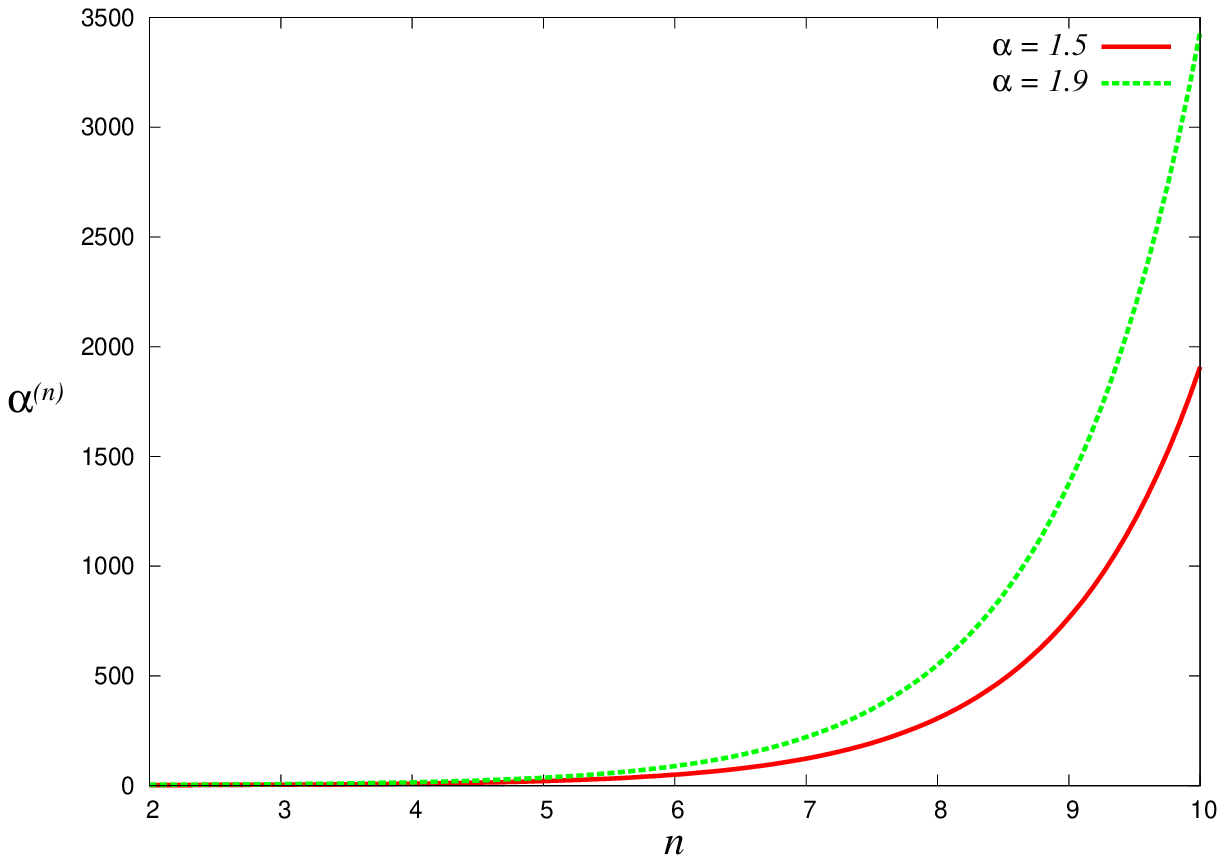}
} 
\caption{\footnotesize 
The $n$-dependence of 
job-offer ratio $\alpha^{(n)}$ 
for the case of $n$-invariant $\Omega=0.4$ (left) 
and $n$-invariant $U=0.4$ (right). 
The behavior of $\alpha^{(n)}$ is completely determined by 
the sign of $\alpha-1$ ($\alpha=1$ is the marginal). 
}
\label{fig:fg2_2}
\end{figure}

From equation (\ref{eq:UOmegaalpha}), one can extract useful information about the $n$-dependence 
of the job-offer ratio on the condition that 
either $U$ or $\Omega$ is $n$-independent. 
We first consider the case in which the labor shortage ratio $\Omega$ is independent of $n$. 
By substituting  (\ref{eq:UOmegaalpha}) with a constant $\Omega$, that is, 
$U^{(k)}=\alpha^{(k)}\Omega +1 -\alpha^{(k)}$ into equation (\ref{eq:alpha_n1}), we have
\begin{equation}
\alpha^{(n)}=\frac{\alpha- (1-\Omega^{n})\prod_{k=0}^{n-1}\alpha^{(k)}}
{1-(1-\Omega^{n})\prod_{k=0}^{n-1}\alpha^{(k)}}.
\end{equation}
Solving the above equation with respect to $\alpha^{(n)}$, we have 
the solution as a function of constant $\Omega$ as 
\begin{equation}
\alpha^{(n)}=\frac{\alpha\Omega^{n}}{1-\alpha (1-\Omega^{n})},\,\,\,
\alpha < 1
\label{eq:invariant_O}
\end{equation}
where we should remember that we defined $\alpha \equiv \alpha^{(0)}$. 
Then, it should be stressed that 
the scale-invariant $\Omega$ is achieved if and only if 
the job-offer ratio $\alpha$ at the initial stage is smaller than unity. 
In other words, 
the scale-invariant $\Omega$ is never achieved for $\alpha >1$. 
From the general result we mentioned in the previous section, for 
the case of $\alpha <1$,  the job-offer ratio $\alpha^{(n)}$ is a monotonically decreasing function of 
$n$. 
Therefore, we ultimately confirm that 
$\lim_{n \to \infty}\alpha^{(n)}=0$ and 
$\lim_{n \to \infty}U^{(n)}=1$. Namely, 
`perfect unemployment state' could be realized at the end of 
our job-hunting process $n \to \infty$. 
In Figure \ref{fig:fg2_2} (left), 
we plot the $\alpha^{(n)}$ as a function of 
$n$ for the initial job-offer ratio  $\alpha=0.5$ and $0.9$. 
\mbox{}
The learning curve for this case is easily obtained as follows. 
Substituting (\ref{eq:invariant_O}) into (\ref{eq:OU}), we have 
$U^{(k)}=
\{1-\alpha (1-\Omega^{k+1})\}/
\{1-\alpha (1-\Omega^{k})\}$. Hence, 
accompanying this result with 
$(1-U)_{n}=1-\prod_{k=0}^{n-1}U^{(k)}$, 
the error-measurement $\varepsilon_{n}$ defined by (\ref{eq:LC}) for the case of $\alpha <1$ is now written as 
\begin{eqnarray}
\varepsilon_{n} & = & 
\alpha - (1-U)_{n} \nonumber \\
\mbox{} & = & 
\alpha -\left\{
1-
\prod_{k=0}^{n-1}
\frac{1-\alpha(1-\Omega^{k+1})}
{1-\alpha (1-\Omega^{k})}
\right\} =  \alpha\Omega^{n}=\alpha\, {\exp}[
-n \log (1/\Omega)]. 
\end{eqnarray}
Therefore, 
the error-measurement $\varepsilon_{n}$ (`learning curve') decays exponentially to zero. 
In other words, 
the cumulative employment rate $(1-U)_{n}$ converges to the theoretical 
upper bound $\alpha\,(<1)$ with an exponential speed. 

On the other hand, 
for the case of $n$-independent 
$U$, by setting $U^{(k)}=U$ in (\ref{eq:alpha_n1}), we immediately obtain 
\begin{equation}
\alpha^{(n)}= 
\frac{\alpha-1 + \prod_{k=0}^{n-1}U^{(k)}}
{\prod_{k=0}^{n-1}U^{(k)}} = 
\frac{\alpha -1 + U^{n}}{U^{n}},\,\,\,\alpha > 1. 
\end{equation}
Here we should notice that 
the scale-invariant $U$ is achieved if and only if 
$\alpha$ satisfies $\alpha >1$. 
Then, the $\alpha^{(n)}$ is a monotonically increasing function 
of $n$ and it reads $\lim_{n \to \infty}\alpha^{(n)}=\infty$ and 
$\lim_{n \to \infty}\Omega^{(n)}=1$. 
This means that 
`perfect labor shortage state' appears 
at the final stage of our job-hunting process. 
When we use the error measurement in 
the collective behavior of students from (\ref{eq:LC}) as  
\begin{equation}
\varepsilon_{n} \equiv 
1-(1-U)_{n}
\end{equation}
for $\alpha >1$, 
the $\varepsilon_{n}$ as a function of 
$n$ could be regarded as a learning curve for this case. 
For the scale-invariant $U$, the curve decays to zero exponentially as 
\begin{equation}
\varepsilon_{n} =U^{n} = {\exp}[-n \log (1/U)]. 
\end{equation}
Therefore, 
the error-measurement decays exponentially to zero, 
in other words, 
the cumulative employment rate $(1-U)_{n}$ converges to unity, 
which means that 
all students can obtain the position asymptotically. 

The above two limiting cases 
seem to be rather special. Thus, one can naturally assume that both $\Omega$ and $U$ 
could be dependent on the job-hunting stage $n$. 
In order to consider such general case somehow or other, we shall revisit our probabilistic labor market \cite{Chen,Chenb}.  
Of course, in macroeconomics (labor science), there exist a lot of effective attempts to discuss 
the macroscopic properties \cite{Roberto,Fagiolo,Neugart} 
including so-called search theory \cite{Lippman,Diamond,Pissarides1985,Pissarides2000}.
However, it should be stressed that the macroscopic approaches lack of their microscopic viewpoint, namely, 
in their arguments, the behaviour of microscopic (heterogeneous) agents such as job-seekers or companies are neglected. 
\section{The model system}
\label{sec:model}
In this section, in order to derive 
the annual $U$-$\Omega$ curve that we have shown in the previous section or 
or $n$-dependence of 
the macroscopic quantities for general scenario, 
we will briefly  review our model system that  
we have defined in previous studies \cite{Chen,Chenb}. 
\subsection{Assumptions}
We first introduce several basic properties which should be satisfied in our probabilistic labor market.  
In our model system, the labors (the job seekers) are restricted to university graduates and the other people 
on the job searching are neglected for simplicity. In our modeling, 
we shall assume 
that the following four simple points (\romannumeral1)--(\romannumeral4) should 
be taken into account to construct the labor markets.
\begin{enumerate}
\item[(\romannumeral1)]
Each company recruits constant numbers of newcomers $v_{k}^{*}\in\mathbb{Z}$ in each year.
\item[(\romannumeral2)] 
If the company takes too much or too less applications which are far beyond or far below the quota, 
the ability of the company to gather applicants in the next business year decreases.
\item[(\romannumeral3)]
Each company is apparently ranked according to various perspectives, and it is available for all students.
\item[(\romannumeral4)]
A diversity of making the decision of labor should be taken into account by means of maximization 
of Shannon's entropy under some relevant constraints.
\end{enumerate}
\subsubsection{Local mismatch measurement}
Considering the condition (\romannumeral2) cited above, we shall define the following {\it local mismatch measurement}: 
\begin{equation}
h_{k}(t) =  \frac{1}{V} |v_{k}^{*}-v_{k}(t)| = \frac{1}{\alpha N} |v_{k}^{*}-v_{k}(t)|
\end{equation}
where  $v_{k}(t)$ denotes the number of students who seek for the position in the company $k$ at the business year $t$. 
Hence, the above local mismatch measurement $h_{k}(t)$ stands for the difference (gap) between the number of applicants $v_{k}(t)$ 
and the quota $v_{k}^{*}$ for each company $k$.  
\subsubsection{Ranking factor}
On the other hand, based on the condition (\romannumeral3), we define the ranking of the company $k$ by  $\epsilon_{k}(>1)$ 
which is independent of the business year $t$. 
Here we assume that the ranking of the company $k$ is higher if the value 
of  $\epsilon_{k}$ is larger. In our agent-based modeling, we simply set the value as  
\begin{equation}
\epsilon_{k} = 1+\frac{k}{K}. 
\label{eq:ranking}
\end{equation}
In such a situation, the company $k=K$ is the highest ranking company and the company $k=1$ is the lowest. 
We should mention that 
in the second term in 
the right hand side of 
(\ref{eq:ranking}), we divided the 
$k$ by the number of companies $K$ so 
as to make the maximum ranking factor $\max_{k}\epsilon_{k}=\epsilon_{K}=2$ 
a size-independent quantity. 
\subsubsection{Energy function: Student's preference of market history or ranking}
With the above definitions, the energy function of our probabilistic labor market for each company $k$ is written explicitly by
\begin{equation}
E(\epsilon_{k}, h_{k};t)  \equiv  -\gamma \log \epsilon_{k} + 
\sum_{l=1}^{\tau}\beta_{k}(t-l) h_{k}(t-l). 
\label{eq:energy}
\end{equation}
When we define  
two $\tau$-dimensional {\it market history vectors}: 
$\mbox{\boldmath $\beta$}_{k} \equiv 
(\beta_{k}(t-1),\cdots,\beta_{k}(t-\tau))$ and 
$\mbox{\boldmath $h$}_{k} \equiv 
(h_{k}(t-1),\cdots,h_{k}(t-\tau))$, 
the second term appearing in the above energy function (\ref{eq:energy})  
is defined by the inner product of these two vectors as 
$(\mbox{\boldmath $\beta$}_{k} \cdot 
\mbox{\boldmath $h$}_{k})$. 
In this paper, we shall simply choose 
a particular market history vector as $\mbox{\boldmath $\beta$}_{k}=(\beta,0,\cdots,0)$ for all $k$.

Based on this energy function, the behavior of students can be considered as follows. 
From the first term of the energy function, students tend to post their applications to companies 
with relatively high ranking. However, from the second term appearing in the energy function, 
we should notice that even if the company $k$  gathers a lot of applicants as over the quota at some year $t$,  
the probability that the company $k$ gathers the applicants at the next business year $t+1$ decreases.   
Therefore, the second term depends on the history of the labor market as a `negative feedback' on the first term. 
The information about the market in the previous business year $h_{k}(t)$ is much more important for the students 
for $\gamma > \beta$, whereas the ranking of companies becomes more essential for $\gamma < \beta$ to decrease the energy function. 

Here we would like to mention 
the relationship between these parameter set $(\gamma, \beta)$ and 
heterogeneity in our labor market. 
Obviously, 
the preferences of companies' ranking and market history 
depend on a private individual  because each students possess 
his/her own strategy to obtain the suitable position. 
In this sense, these parameters should be selected as 
$(\gamma_{i}, \beta_{i})$ for 
$i=1,\cdots, N$. 
However, this type of choice might cause tremendous difficulties in 
our modeling because we should calibrate $2N$ parameters from 
the empirical evidence. 
Of course, 
one can assume that 
$\gamma$ and $\beta$ are generated by some 
parametric distributions 
$P(\gamma)$ and $P(\beta)$ which 
contain a finite number of parameters. 
However, for this choice also, 
we need to determine the parameters from limited empirical data and 
it is not so easy for us to carry out. 
Therefore, in this paper, we simply assume that 
the parameters are independent of student. 
As we see in the next section, 
heterogeneous properties (diversities) in our model system 
are induced by means of the Jaynes-Shannon's MaxEnt principle. 
\subsection{Jaynes-Shannon's MaxEnt principle}
In the previous sections, we introduced the energy function (\ref{eq:energy}) to be 
minimized for each student to choose the company to apply for. 
However, minimizing the energy function leads to 
a unique solution without any diversity. 
The lack of diversity for the students' choices 
sometimes causes the tremendous global mismatch. 
In order to introduce the diversity into 
the system, we utilize the so-called 
{\it Jaynes-Shannon's MaxEnt principle} \cite{Jaynes}. 
In following, we derive the aggregation probability for the students by means of 
the maximum entropy principle. 

As we mentioned,  the decision-makings of labors that make up the labor market should be `diverse' 
from the assumption (\romannumeral4). 
In order to quantify the diversity, we maximize the following Shannon's entropy: 
\begin{equation}
H = -\sum_{k=1}^{K}P_{k}(t) \log P_{k}(t)
\end{equation}
under the normalization condition of the probability $P_{k}(t)$, 
namely, 
\begin{equation}
\sum_{k=1}^{K}P_{k}(t) = 1
\end{equation}
and the conservation of energy 
\begin{equation}
E = \sum_{k=1}^{K}P_{k}(t)E(\epsilon_{k}, h_{k};t)
\end{equation}
in the each time (annual). 
Then, we consider the following 
variational problem described by the functional \cite{Jaynes} 
concerning $P_{k}(t)$: 
\begin{equation}
f[P_{k}(t)] = 
-\sum_{k=1}^{K}P_{k}(t) \log P_{k}(t) + 
\lambda_{1} 
\left\{1-\sum_{k=1}^{K}P_{k}(t)
\right\}+ 
\lambda_{2} 
\left\{E - \sum_{k=1}^{K}P_{k}(t)E(\epsilon_{k}, h_{k};t)
\right\}
\end{equation}
where $\lambda_{1}$ and $\lambda_{2}$ are Lagrange multipliers. 
By solving this variational problem with respect to $P_{k}(t)$ so as to satisfy 
\begin{equation}
\frac{\delta f[P_{k}(t)]}{\delta P_{k}(t)} = 
\frac{\delta f[P_{k}(t)]}{\delta \lambda_{1}} = 
\frac{\delta f[P_{k}(t)]}{\delta \lambda_{2}}=0, 
\end{equation}
the probability $P_{k}(t)$ that the company $k$ gathers the candidates at time $t$ is given by 
\begin{equation}
P_{k}(t)= 
\frac{{\exp}
\left[
\gamma \log (1+\frac{k}{K}) -\beta |v_{k}^{*}-v_{k}(t-1)|
\right]}
{
\sum_{k=1}^{K}
{\exp}
\left[
\gamma \log (1+\frac{k}{K}) -\beta |v_{k}^{*}-v_{k}(t-1)|
\right]}
\label{eq:pkt}
\end{equation}
where we chose 
${\rm e}^{-1-\lambda_{1}}=\{
\sum_{k=1}^{K}
{\exp}[
\gamma \log (1+\frac{k}{K}) -\beta |v_{k}^{*}-v_{k}(t-1)|
]
\}^{-1}$ and 
set $\lambda_{2}=1$ (unit `temperature' to control the diversity of the system) for simplicity. 
Therefore, 
for each time step $t$, 
each student posts his/her application letter to 
the company $k$ according to the above 
aggregation probability (\ref{eq:pkt}). 

It is worth while for us to mention that 
the probability described by (\ref{eq:pkt}) has a quite similar formula to   
the so-called {\it multinomial (non-linear) logit model} or {\it softmax regression} in the literature of statistics 
(see {\it e.g.} \cite{McFadden,Train}). 
Actually, when we define the so-called softmax function as 
\begin{equation}
{\rm softmax}(k, x_{1},x_{2},\cdots, x_{K}) = 
\frac{
{\exp} (x_{k})}
{\sum_{i=1}^{K} {\exp}(x_{i})}, 
\end{equation}
the aggregation probability $P_{k}(t)$ is written by 
\begin{equation}
P_{k}(t) = {\rm softmax}(k,-E_{1}(t),-E_{2}(t),\cdots,-E_{K}(t)). 
\end{equation}
Henceforth, 
the actions of students to post their application letters 
are governed by the aggregation probability (\ref{eq:pkt}). 
To monitor 
their `microscopic' actions, we introduce several microscopic and macroscopic quantities 
in the following sections. 
\subsubsection{Microscopic variables}
Here we define the microscopic variable $s_{ik}(t)$ which is used to distinguish whether the student  $i$ receives 
an acceptance from the company $k$ in the business year $t$ or not. 
Namely, 
$s_{ik}(t)$=1 means that the student  $i$ receives an acceptance from the company 
$k$ in the business year $t$, and $s_{ik}(t)$=0 means that he/she does not. 
In particular, the microscopic variable $a_{ik}(t)$ is used to distinguish whether 
the student $i$ decides to post  his/her application letter to the company $k$ at time $t$ or not as 
\begin{equation}
a_{ik}(t) = \left\{ \begin{array}{ll}
 1 & \textrm{(with $P_{k}(t)$)}\\
 0 & \textrm{(with $1- P_{k}(t)$)}\\
  \end{array} \right.
\label{eq:a_{ik}}
\end{equation}
Then, we should notice that the total number of acceptances 
for the student $i$ is defined by 
\begin{equation}
s_{i}(t) = \sum_{k=1}^{K}s_{ik}(t)
\end{equation}
with
\begin{equation}
\begin{split}
P(s_{ik}(t) = 1 | a_{ik}(t))=
\Theta(v_{k}^{*}-v_{k}(t)) \delta_{a_{ik}(t),1} + \frac{v_{k}^{*} }{v_{k}(t)}\, \Theta (v_{k}(t)-v_{k}^{*}) \delta_{a_{ik}(t),1}
\label{eq:Psik}
\end{split}
\end{equation}
and $P(s_{ik}(t) = 0 | a_{ik}(t) )= 1- P(s_{ik}(t) = 1 | a_{ik}(t))$, 
where  $\delta _{x,y}$ stands for the Kronecker's delta, 
and $\Theta(x)$ denotes the conventional step function defined by 
\begin{equation}
\Theta (x) = 
\left\{
\begin{array}{cc}
1 & (x \geq 1) \\
0 & (x <0)
\end{array}
\right.
\end{equation}
Thus, the first term of equation (\ref{eq:Psik}) means that the $s_{ik}(t)$ 
takes $1$ with unit probability when the student $i$ posts the application to the company $k$ 
and the total number of applications the company $k$ gathered does not exceed the quota $v_{k}^{*}$. 
On the other hand, the second term means that the 
$s_{ik}(t)$ takes $1$ with probability $v_{k}^{*}/v_{k}(t)$ even if $v_{k}(t)>v_{k}^{*}$ holds. 
In other words, for $v_{k}(t)>v_{k}^{*}$, the informally accepted $v_{k}^{*}$ students are randomly selected from 
$v_{k}(t)$ candidates. 

In this paper, 
we use the above `random selection' by companies, 
however, it is possible for us to extend the selection 
process by companies to much more realistic version. 
For instance, 
in the reference \cite{Chen2}, 
we considered a scenario in which 
each company that obtained application letters over the quota 
selects the candidates according to their scores (school records). 
 \section{The results}
 \label{sec:result}
 In this section, we carried out agent-based simulations 
 based on our probabilistic model 
 described by  (\ref{eq:pkt})(\ref{eq:Psik}) and (\ref{eq:a_{ik}}). 
 Before we simulate the artificial labor market, 
 we should determine (estimate) several parameters such as 
 the degree of 
 ranking preference $\gamma$, 
 market history preference $\beta$ and 
 average number of application letters $a$ 
 from the available empirical data. 
 \subsection{Model calibration and theoretical $U$-$\Omega$ curve}
 To determine the parameters appearing in the system, we focus on the following 
 unemployment rate as an order parameters. 
\begin{equation}
U \equiv U_{0}=U^{(0)}=
\lim_{T\to \infty}
\frac{1}{T} \sum_{t=0}^{T-1}U_{t}, \,\,\,\,\,
U_{t}  =  \frac{1}{N}\sum_{i=1}^{N}
\delta_{s_{i}(t),0}.
\label{eq:orderP} 
\end{equation}
The labor shortage ratio $\Omega \equiv \Omega_{0}=\Omega^{(0)}$ is evaluated by 
inserting the above result (\ref{eq:orderP}) into equation (\ref{eq:UOmegaalpha}). 
As we mentioned, in order to draw the $U$-$\Omega$ curve, we need information about the system parameters 
such as strength of ranking preference $\gamma$ and market history $\beta$. 
Here we fix $\beta=1$ and look for the best possible $\gamma$ as a solution of 
$U_{\rm empirical}=U(\gamma)$, 
namely, 
\begin{equation} 
\gamma = U^{-1}(U_{\rm empirical}),
\label{eq:inverse}
\end{equation}
where $U_{\rm empirical}$ denotes the empirical unemployment rate 
for a given job-offer ratio $\alpha \equiv \alpha^{(0)}$, which is opened for the public  by MEXT $\&$ MHLW \cite{mext} in every October 
(see Figure \ref{fig:fg2}(left)). 
In the reference \cite{Chen,Chenb}, 
we showed that the cumulative employment rate $1-U=(1-U)_{0}$ 
monotonically decreases as $\gamma$ increases. It means that 
there exists a unique solution of (\ref{eq:inverse}) for a given empirical 
data $U_{\rm empirical}$. 
In our simulations, 
we set $N=2000, K=100$ and 
$n_{k}^{*}=\eta, \forall_{k}$, and it turns out to be 
\begin{equation}
\eta = \frac{N}{K}\alpha=20\alpha
\end{equation}
for a given $\alpha$. 
Hence, we can control the $\eta$ to realize any positive value of $\alpha$. 

\begin{figure}[!htb]
\resizebox{\textwidth}{!}{%
  \includegraphics{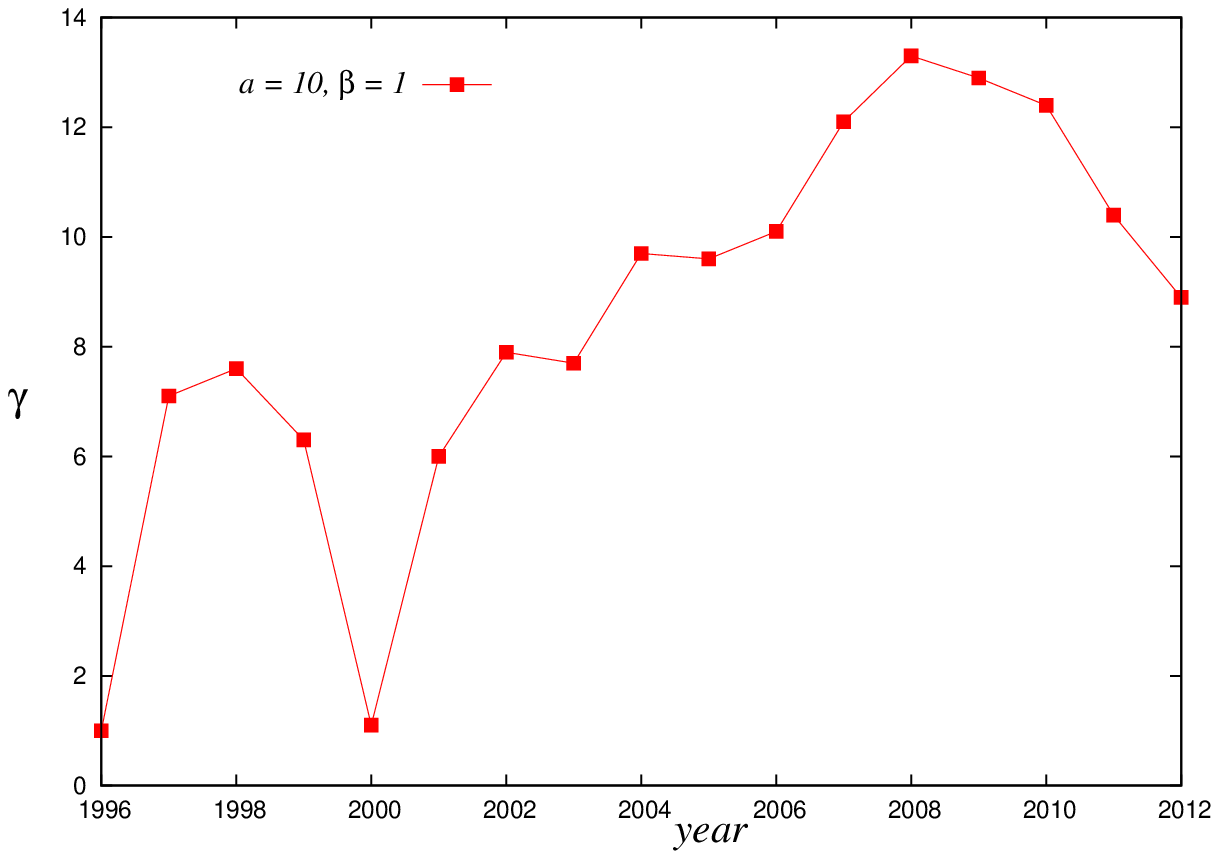}
   \includegraphics{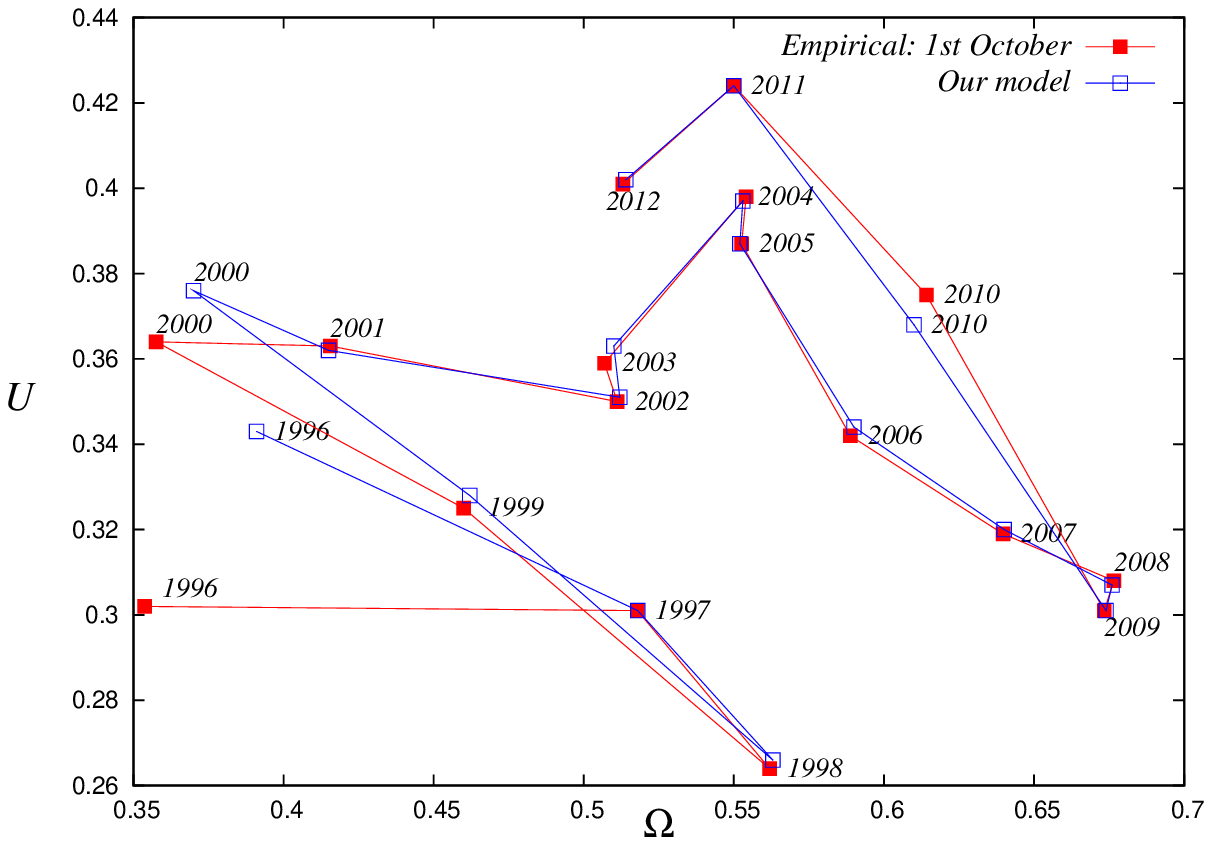}
}
\caption{\footnotesize 
Time dependence of ranking preference $\gamma$ (left). 
By using this panel accompanying with the right panel in Figure \ref{fig:fg2} (right), 
one can account for the appearance of global mismatch.  
The right panel shows
the empirical and theoretical $U$-$\Omega$ curves. 
We clearly find that the large $\gamma$ apparently pushes 
the $U$-$\Omega$ curve toward the upper right direction where 
the global mismatch between the students and companies is large. 
}
\label{fig:fg3}
\end{figure}

We plot the resulting $\gamma$ and $U$-$\Omega$ curve in Figure \ref{fig:fg3}. 
From this figure, we find that the large $\gamma$ apparently pushes 
the $U$-$\Omega$ curve toward the upper right direction where 
the global mismatch between the students and companies is extremely large. 
The gap between the theoretical and empirical $U$-$\Omega$ curve comes from 
the uncertainties in the calibration of average number of 
application letters $a$. Obviously, it is observable quantity and 
a large scale survey tells us the precise statistics. 
However, up to know, there is no authorized reliable statistics 
although several private researches provided the data. 
Hence, we simply chose the value as $a=10$ in our simulation.

We should bear in mind that 
the parameter $\gamma$ is not observable in principle. 
However, our approach enables us to 
mention the annual strength of students' ranking preferences $\gamma$ against 
that of the market history $\beta$.  
In 2010 when the global mismatch was the largest, 
the point $(\Omega,U)$ is located at the upper right position in the 
$U$-$\Omega$ curve. 
However, 
the corresponding $\gamma$ does not take the peak in the annual $\gamma$ trends 
and the peak appeared in 2008. 
At a glance, it seems to be inconsistent, however from 
Figure \ref{fig:fg2} (right), we find that 
the job-offer ratio $\alpha$ in 2010 is larger than that in 2008. 
This implies that the market in 2010 was more `seller's market' than in 2008, 
and the market supplied a lot of positions to job-seeking students. 
Actually, unemployment rate $U$ in 2008 was relatively low (see Figure \ref{fig:fg2} (left)). 
On the other hand, in 2010, 
the unemployment rate increases by $0.6$ from 2009 due to 
the decreases of $\alpha$ although 
the strength of students' ranking preferences was getting smaller during 2009-2010. 
As the result, the point $(\Omega,U)$ moved toward the upper right position in the $U$-$\Omega$ curve (see Figure \ref{fig:fg3} (right)), 
which resulted in the worst global mismatch between students and companies in 2010. 
\subsection{Analysis for the general case}
As we already mentioned before, 
both $U$ and $\Omega$ are not scale-invariant in general. 
Therefore, we should investigate the flow of relevant quantities $(\alpha^{(n)},U^{(n)},\Omega^{(n)})$ by means of 
our probabilistic labor market \cite{Chen,Chenb}. 
After several complicated algebra, we obtain the following coupled `dynamical' equations. 
\begin{eqnarray}
P_{k}(t) & = &  
\frac{{\exp}
\left[
\gamma \log (1+\frac{k}{K}) -\beta |v_{k}^{*}-v_{k}(t-1)|
\right]}
{
\sum_{k=1}^{K}
{\exp}
\left[
\gamma \log (1+\frac{k}{K}) -\beta |v_{k}^{*}-v_{k}(t-1)|
\right]}
\label{eq:Pk_t} \\
\alpha^{(t)} & = &  \frac{\sum_{k=1}^{K}
(v_{k}^{*}(t)-m_{k}(t))
\delta_{v_{k}^{*}(t),m_{k}(t)}}
{
\sum_{i=1}^{N}
\prod_{l \leq t}
\delta_{s_{i}(l),0}} \\
U^{(t)}  & = &  1-
\frac{\sum_{i=1}^{N}
\prod_{l \leq t}
\delta_{s_{i}(l),0}}
{\sum_{i=1}^{N}
\prod_{l \leq t-1}
\delta_{s_{i}(l-1),0}}
\label{eq:U_t}  \\
\Omega^{(t)} & = &  
\frac{U^{(t)}+\alpha^{(t-1)}-1}
{\alpha^{(t-1)}} \\
(1-U)_{n} & = &  \frac{\sum_{i=1}^{N}
\prod_{l \leq t}
\delta_{s_{i}(l),0}}{N}
\label{eq:Omega_t}
\end{eqnarray}
We carry out computer simulations 
based on the above equations (\ref{eq:Pk_t})-(\ref{eq:Omega_t}). 
\subsubsection{Collective behavior of zero-intelligence students}
In our simulation, we calibrate the parameters 
$\gamma, \beta$ by making use of 
the empirical data which is opened to the public 
in every October. 
After the calibration, we do not change the variables. 
In this sense, the students do not `learn' at all from 
the past markets (stages). 
In real labor market for university graduates, 
students who failed to get the position at $n=0$ 
might change their strategies. 
For instance, they might decreases the strength of 
ranking preference $\gamma$ to look for much broadly including 
small enterprises. 
However, we do not take into account such `intelligence' of the students and 
we investigate to what extent the collective behaviour of such `zero-intelligence' students 
can reduce the global mismatch. 

\begin{figure}[!htb]
\resizebox{\textwidth}{!}{%
 \includegraphics{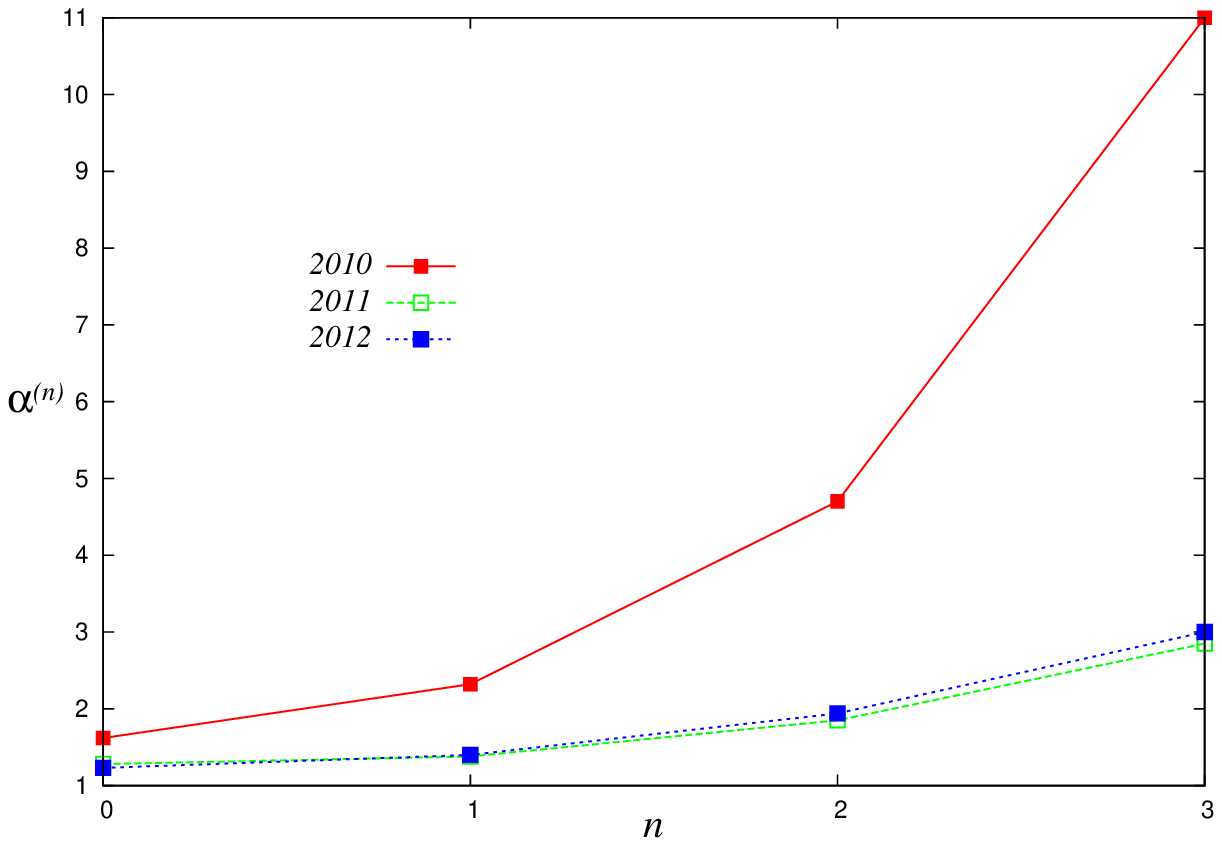}
  \includegraphics{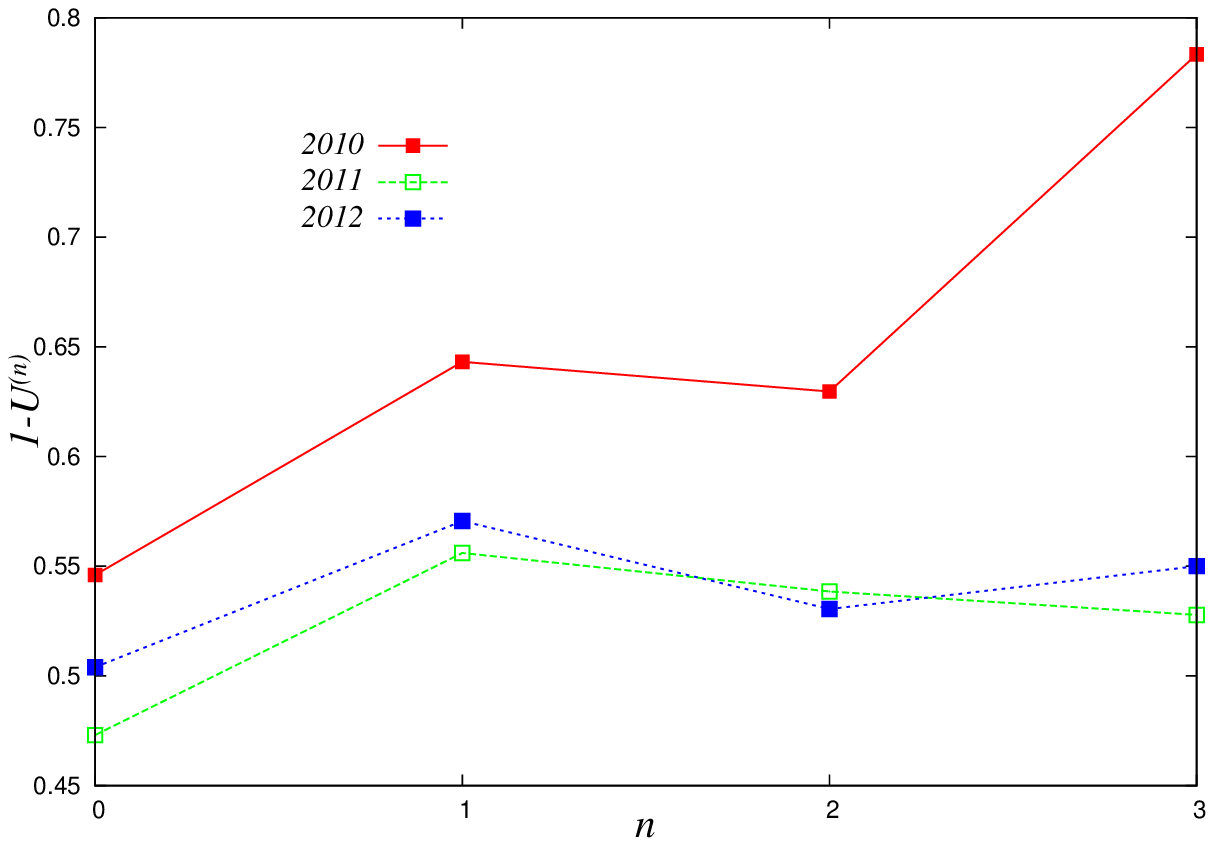}
} \\
\resizebox{\textwidth}{!}{%
  \includegraphics{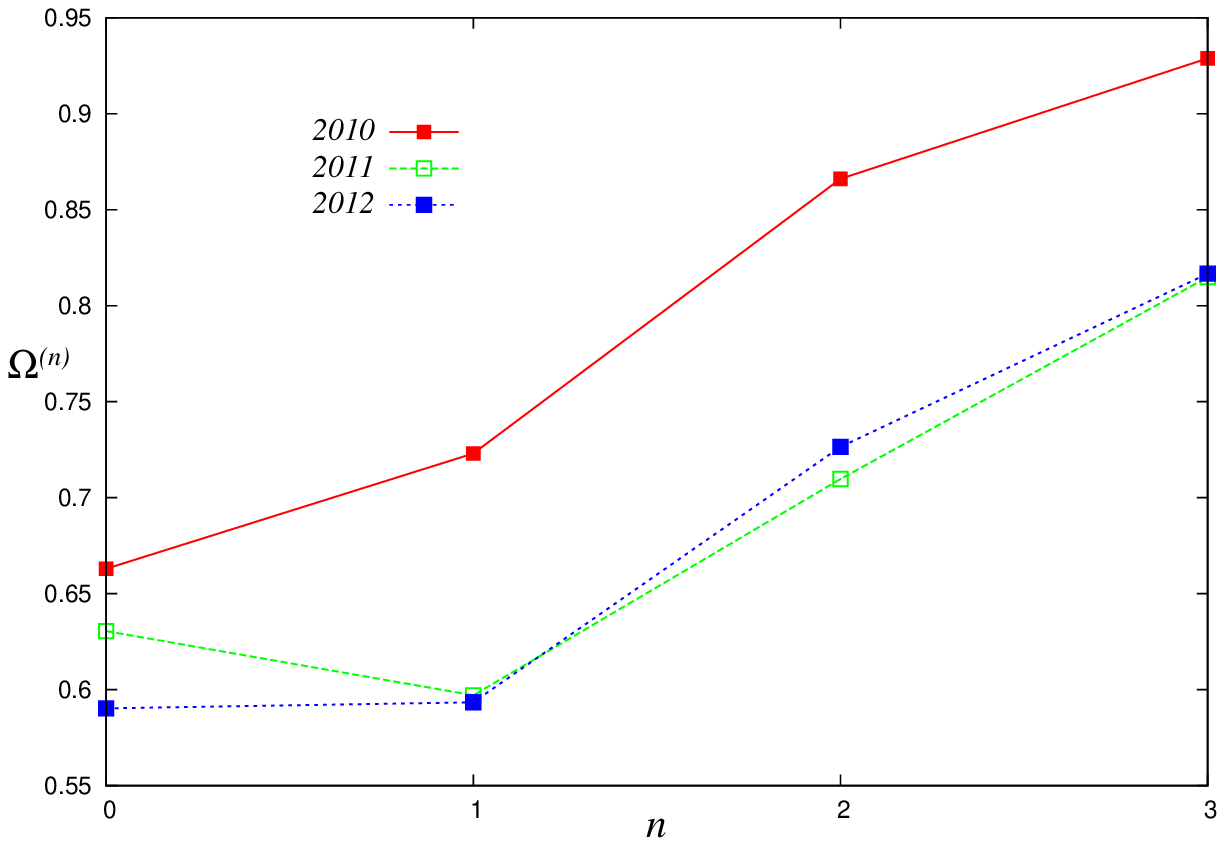}
    \includegraphics{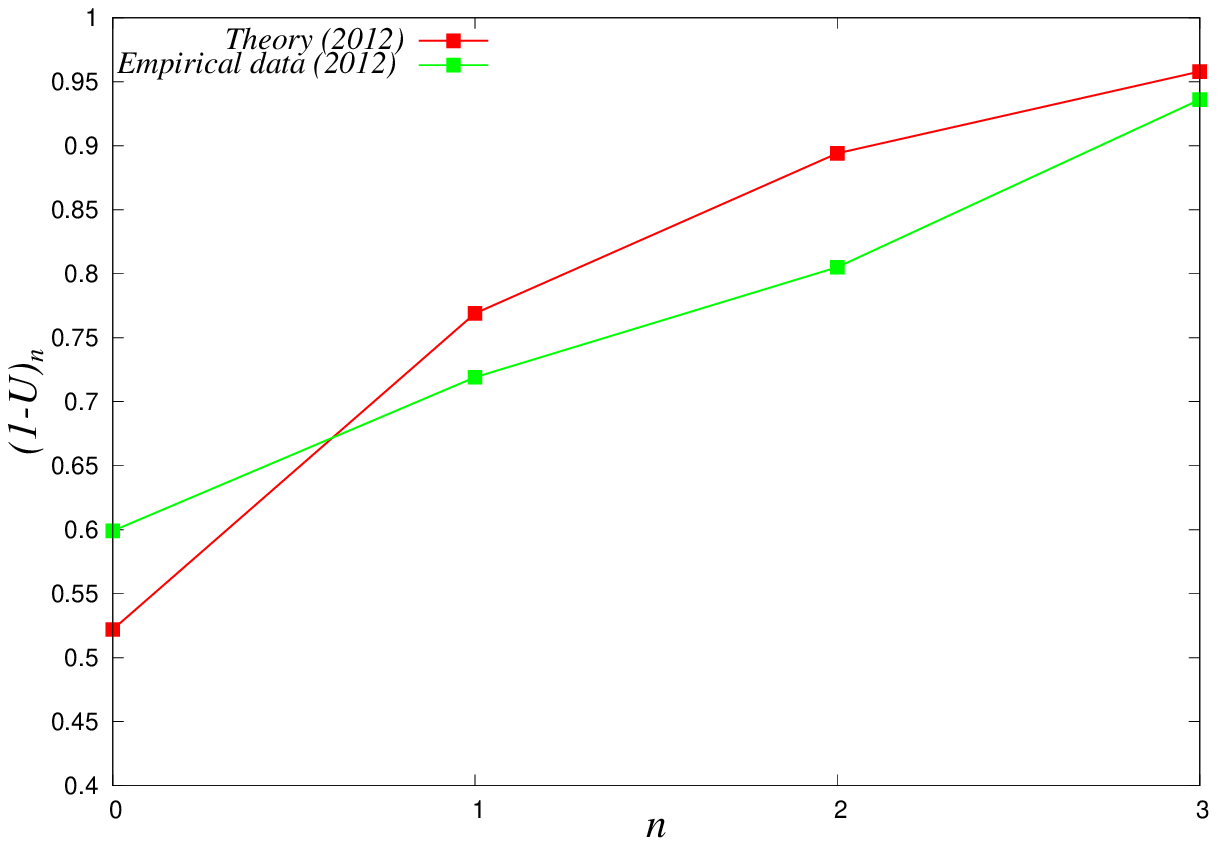}
  }
\caption{\footnotesize 
From the upper left to the lower right, 
$\alpha^{(n)}$ and $1-U^{(n)},\Omega^{(n)}$ in the last three years, 
and the cumulative employment rate $(1-U)_{n}$ as a function of $n$ in 2012 
are shown. 
}
\label{fig:fg4}
\end{figure}

In Figure \ref{fig:fg4}, 
we plot $\alpha^{(n)}$ and $1-U^{(n)},\Omega^{(n)}$ in the last three years, 
and the cumulative employment rate $(1-U)_{n}$ as a function of $n$ in 2012. 
From these panels, we clearly find that 
the relevant quantities 
$(\alpha^{(n)},U^{(n)},\Omega^{(n)})$ 
are not scale-invariant at all and 
each of them possesses a rather complicated $n$-dependence. 
It should be stressed that 
the empirical data set  for $(\alpha^{(n)},U^{(n)},\Omega^{(n)})$ 
is not available (has never been released to the public) and we just have the data for the cumulative 
employment rate $(1-U)_{n}$ instead of $U^{(n)}$ as shown in 
Figure \ref{fig:fg2}. 
In this sense, our result might be addressed as the `first attempt' to 
show the stage $n$-dependence of the quantities $(\alpha^{(n)},U^{(n)},\Omega^{(n)})$ explicitly 
in Japanese labor market. 

In the lower right panel, we show 
the cumulative employment rate $(1-U)_{n}$ with the counter empirical plot. 
We are confirmed that the theoretical result is remarkably improved from 
the simple observation $(1-U)_{n}=1-U^{n}$ as shown in 
Figure \ref{fig:fg2} (right).  However, 
there still exists a gap between empirical and theoretical results. 
The gap might come from wrong calibration of system parameters such as 
the average number of application letters $a$ for a student {\it etc.} 
(In the above simulations, we set $a=10$). 

\begin{figure}[!htb]
\resizebox{\textwidth}{!}{%
  \includegraphics{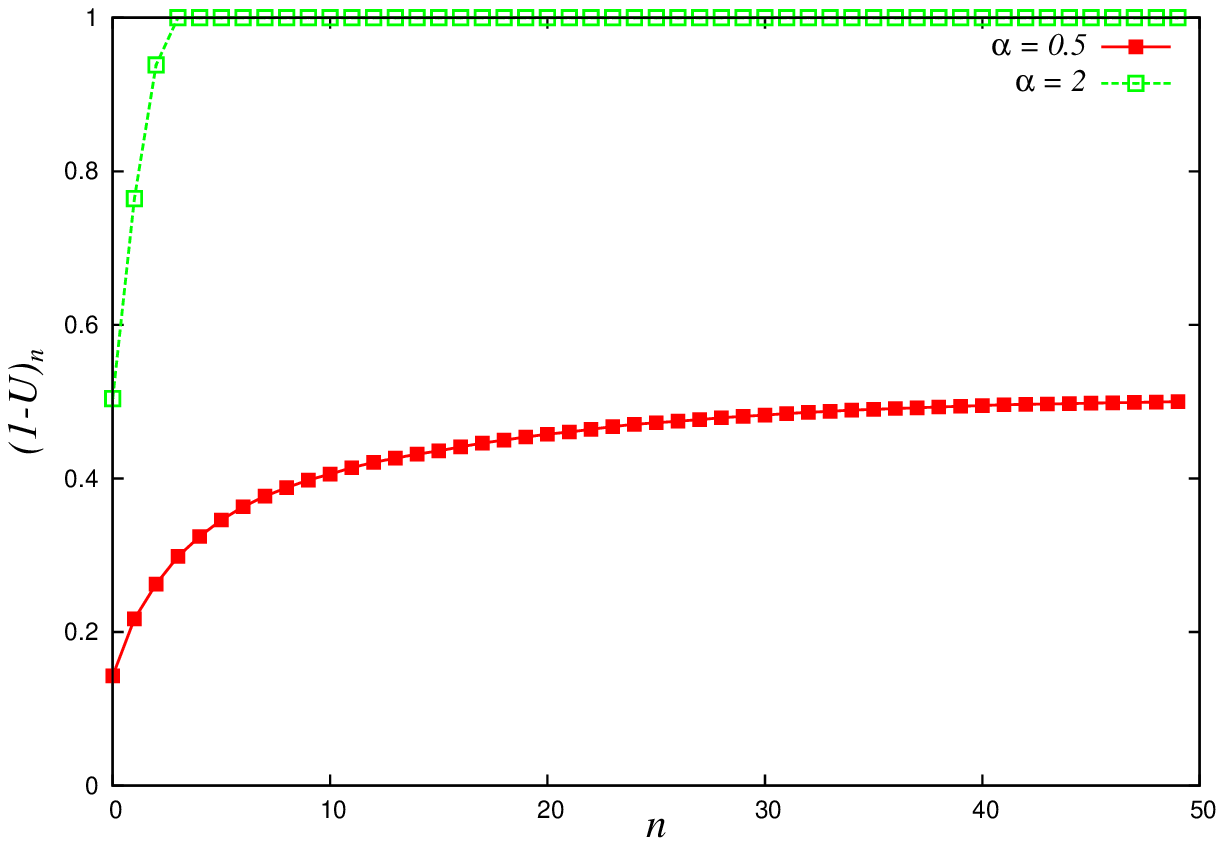}
 \includegraphics{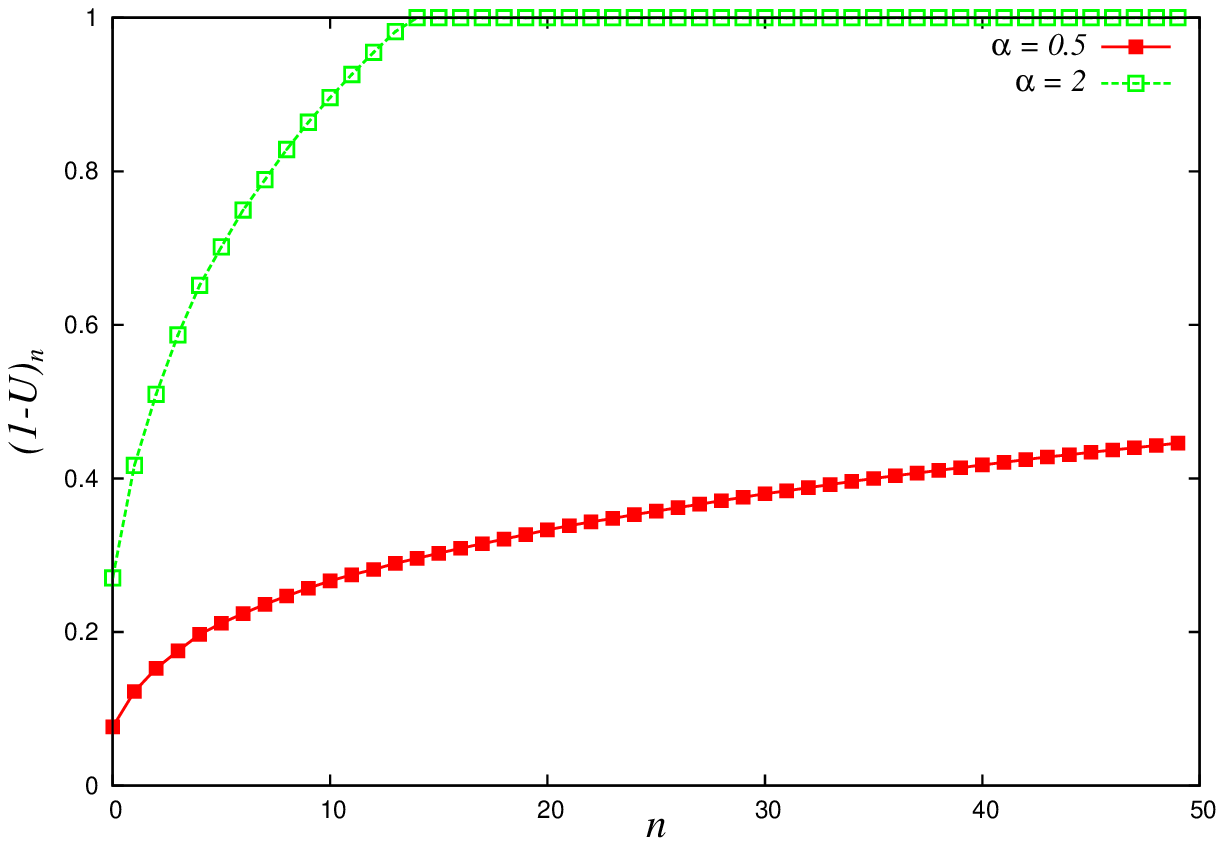}
}
\caption{\footnotesize 
The learning curves as the cumulative employment rate $(1-U)_{n}$ for the case of 
$a=5$ (left) and $a=10$ (right).  We set the system parameters 
simply as $\gamma=\beta=1$. 
}
\label{fig:fg2222}
\end{figure}

Finally, in Figure \ref{fig:fg2222}, 
we show the learning curves as the cumulative employment rate $(1-U)_{n}$ for the case of 
$a=5$ (left) and $a=10$ (right). 
In these panels, we examine 
for $\alpha =0.5$ and $2$. 
From this figure, we clearly observe 
that 
the $(1-U)_{n}$ converges to $1$ for $\alpha=2 \,(>1)$ and 
$\alpha$ for $\alpha=0.5\,(<1)$ as we have discussed in the previous sections. 
Here we should focus on the speed of convergence. 
From both panels, we find that 
the speed of the convergence of $(1-U)_{n}$ to the theoretical upper bound unity for  the case of $\alpha >1$ is much faster than 
that of the case for $\alpha <1$ (the theoretical upper  bound is $\alpha$). 
This fact might imply that 
the job-matching between students and companies in `buyer's market' is much harder than that in `seller's market' especially in the large $n$ regime. 
It is quite reasonable because $\alpha^{(n)}$ decays to zero for $\alpha <1$. 
Another remarkable finding is about average number of application letters, and actually 
we find that the increase of $a$ causes tremendous speed down in the convergence. 
\section{Summary}
Inspired by the unsupervised  learning or self-organization in the machine learning context, 
we drew `learning curve' for the collective 
behavior of job-seeking `zero-intelligence' labors in 
successive job-hunting process. 
We discussed the scale invariance of the macroscopic quantities in 
probabilistic labor markets.  We obtained 
the flow $(\alpha^{(n)},U^{(n)},\Omega^{(n)})$ 
with the assistance of computer simulations. 
Obviously, students in real labor markets are much more `intelligent'  than the agents in our model system. 
In this sense, we should modify our model by taking into account the students' tendency of job-hunting process. 
Especially, `intelligent' learning algorithm of system parameters such as $\gamma$ or $\beta$ might play important role in 
the next direction of our studies. These parameters might be dependent on 
the business cycle in the society and in this sense, 
the labor market should be coupled with the other kinds of markets such as 
financial market or housing market. These issues should be addressed as our future problems.

\section*{Acknowledgement}
This work was financially supported by 
Grant-in-Aid for Scientific Research (C) 
of Japan Society for 
the Promotion of Science, No. 22500195. 
We acknowledge {\it Recruit Works Institute}  
for allowing us to use their very recent survey data \cite{Recruit}.  
One of the authors (JI) thanks 
Yuji Aruka, Yoshi Fujiwara and Hideaki Aoyama 
for fruitful discussion and for encouraging us to submit our new result 
to {\it Evolutionary and Institutional Economics Review}. 

\end{document}